\documentclass[preprint,12pt,authoryear]{elsarticle}

\usepackage{amssymb, amsmath}
\usepackage{graphicx}
\usepackage{footnote}
\usepackage{xcolor}
\usepackage[hyphens]{url}
\usepackage[hidelinks]{hyperref}
\hypersetup{breaklinks=true}

\begin{document}

\begin{frontmatter}



\title{Of Degens and Defrauders: Using Open-Source Investigative Tools to Investigate Decentralized Finance Frauds and Money Laundering}


\author[inst1,inst2]{Arianna Trozze}

\affiliation[inst1]{organization={Department of Computer Science, University College London},
            addressline={Gower Street}, 
            city={London},
            postcode={WC1E 6EA}, 
            country={United Kingdom}}

\author[inst2]{Toby Davies}
\author[inst2,inst3]{Bennett Kleinberg}

\affiliation[inst2]{organization={Dawes Centre for Future Crime, University College London},
            addressline={35 Tavistock Square}, 
            city={London},
            postcode={WC1H 9EZ}, 
            country={United Kingdom}}

\affiliation[inst3]{organization={Department of Methodology and Statistics, Tilburg University},
            addressline={Warandelaan 2}, 
            city={Tilburg},
            postcode={5037 AB}, 
            country={Netherlands}}

\begin{abstract}
Fraud across the decentralized finance (DeFi) ecosystem is growing, with victims losing billions to DeFi scams every year. However, there is a disconnect between the reported value of these scams and associated legal prosecutions. We use open-source investigative tools to (1) investigate potential frauds involving Ethereum tokens using on-chain data and token smart contract analysis, and (2) investigate the ways proceeds from these scams were subsequently laundered. The analysis enabled us to (1) uncover transaction-based evidence of several rug pull and pump-and-dump schemes, and (2) identify their perpetrators' money laundering tactics and cash-out methods. The rug pulls were less sophisticated than anticipated, money laundering techniques were also rudimentary and many funds ended up at centralized exchanges. This study demonstrates how open-source investigative tools can extract transaction-based evidence that could be used in a court of law to prosecute DeFi frauds. Additionally, we investigate how these funds are subsequently laundered.
\end{abstract}


\end{frontmatter}


\section{Introduction}
\label{sec:introduction}


Decentralized finance (DeFi) refers to a system of financial products and services created by smart contracts on blockchains like Ethereum. Fraud across the DeFi ecosystem is a growing concern, with victims losing an estimated \$7.8 billion in cryptocurrency in 2021 to various types of DeFi scams. DeFi-based money laundering from cybercrimes also increased by an estimated 1,964\% from 2020 to 2021 \citep{chainalysis_2022_2022}. Despite this reported growth, associated enforcement actions remain minimal, with only 50 cases having been completed specifically involving DeFi tokens in the United States as of the end of November 2022 \citep{blockchain_association_ba_nodate}; many of these involved Initial Coin Offering (ICO) scams completed prior to DeFi's more widespread adoption. While responsibility for DeFi's oversight remains disputed among enforcement agencies, so far, the U.S. Securities and Exchange Commission (SEC) has asserted its authority and argued in many cases that DeFi tokens constitute securities (see \citep{lbry}).

Existing literature \citep{wang_ponzi_2021, hu_transaction-based_2021, fan_-spsd_2021, xia_demystifying_2021, mazorra_not_2022} focuses on \textit{detecting} various categories of DeFi-based securities violations, such as Ponzi schemes and rug pulls (a type of exit scam). However, all of these studies except that by \citet{xia_demystifying_2021} primarily present results at an aggregate level (and even \citet{xia_demystifying_2021} only explore such violations on a single platform). While this is useful to characterize the landscape of DeFi fraud, and the extent to which these scams are detectable, there is a disconnect between the scale of the frauds these papers detail and prosecutions which address them. 

Our research therefore focuses on using open-source investigative tools to extract evidence of Ethereum-based DeFi frauds that could be used in prosecuting them. We use these tools to (1) investigate potential frauds using on-chain data and token smart contract analysis, and (2) investigate the ways that proceeds from these scams were ultimately laundered. We extract transaction-based evidence which could potentially be used in a court of law. The on-chain evidence we extract also offers insight into how DeFi frauds are committed on Ethereum. In addition to determining how the frauds were executed we also investigate how the proceeds of these schemes were subsequently laundered.

Our research questions are the following:
\begin{enumerate}
    \item What evidence of Ethereum scams can we glean from open-source investigative tools that could be used in prosecuting them?
    \item What can open-source investigative tools tell us about how Ethereum DeFi-based frauds are committed?
    \item What can open-source investigative tools tell us about how perpetrators launder the proceeds of Ethereum DeFi-based frauds?
\end{enumerate}

\noindent This study makes the following contributions to research on this topic:
\begin{itemize}
    \item We demonstrate how open-source investigative tools can be used to extract transaction-based evidence of Ethereum-based frauds that could be used in a court of law to prosecute such scams. 
    \item In addition to determining how the Ethereum-based DeFi frauds were carried out, we investigate how these funds are subsequently laundered.
    \item Finally, we conduct these on-chain investigations more systematically, providing a blueprint for investigators or researchers to use open-source investigative tools to conduct granular DeFi fraud investigations.
\end{itemize}
Against this background, this article begins with an overview of Ethereum and DeFi \footnote{\footnotesize In this research, we focus only on Ethereum-based DeFi, though we acknowledge that DeFi applications exist on manifold blockchains. In this article, where we refer to DeFi, we mean Ethereum-based DeFi.}, followed by an exploration of DeFi fraud and money laundering. We then discuss prior work on detecting DeFi fraud, with an emphasis on rug pulls (a commonly-committed DeFi fraud). We then outline our investigative methods, present the results of our investigations, and discuss our findings and their wider implications. 

\begin{figure*}[h!]
    \centering
    \includegraphics[width=14cm]{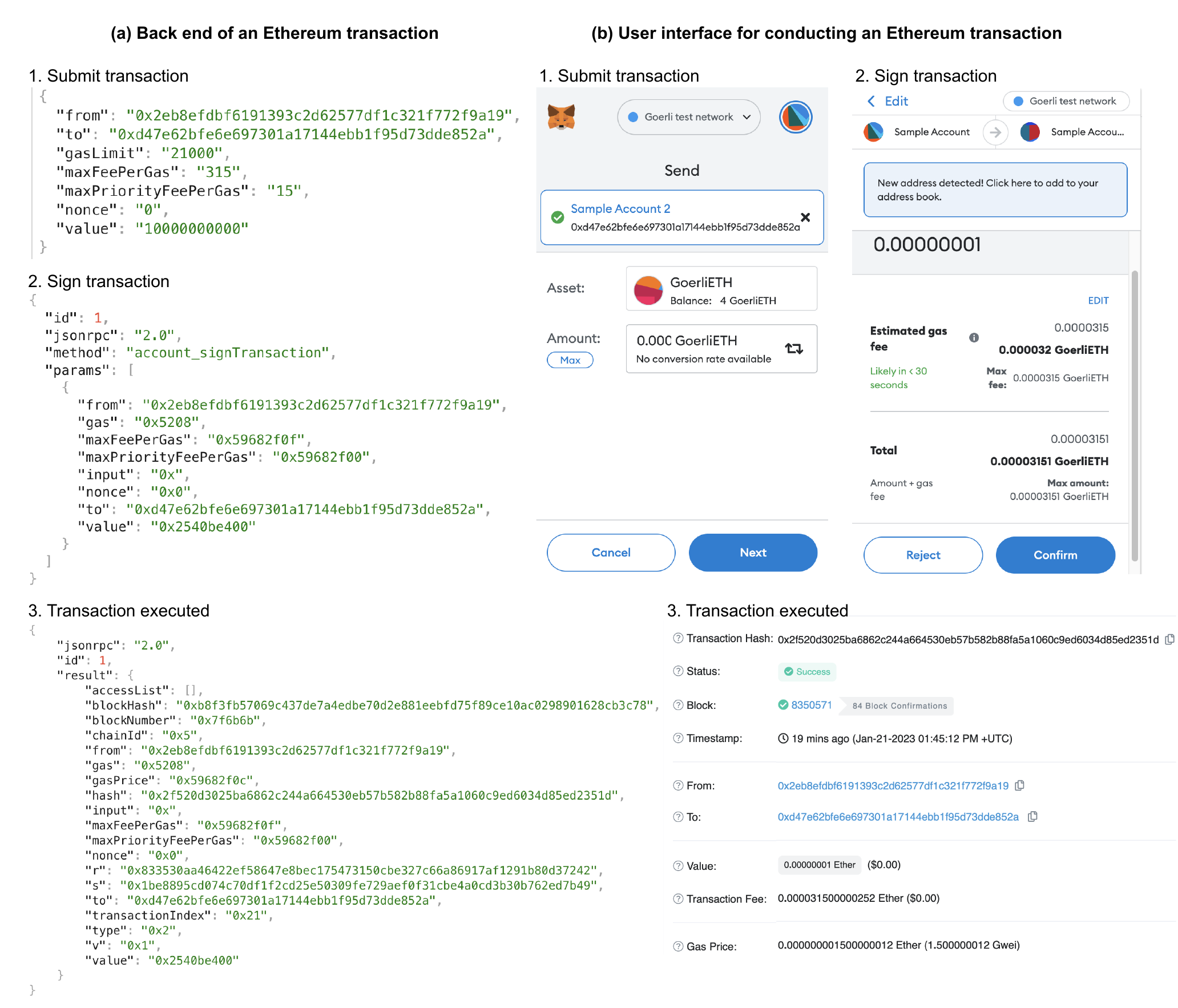}
    \caption{Ethereum transactions. Back end and front end of conducing example Ethereum transaction on the Goerli test network.\\
    (a) Back-end of transaction object submitted to an Ethereum client such as Geth.\\ 1. Transaction oject submitting transaction.\\ 2. JSON-RPC call to sign transaction with user's private key.\\ 3. JSON response showing completed transaction.\\
    (b) User interface for conducting an Ethereum transaction using a Metamask software wallet.\\ 1. Submit transaction.\\ 2. Sign transaction by “confirming" it.\\ 3. Transaction shown as being completed on Etherscan blockchain explorer.}
    \label{fig1}
\end{figure*}

\subsection{Introduction to Ethereum and Decentralized Finance}
In 2008, a pseudonymous developer going by the name Satoshi Nakamoto envisioned a novel financial system, whereby participants could transact with one another in a peer-to-peer manner, rather than through a centralized authority \citep{nakamoto_bitcoin:_2008}. Transactions would be recorded in a distributed ledger (called a blockchain) through an innovative combination of existing cryptographic primitives \citep{narayanan_blockchains_2018}. In 2014, a group of developers extended this idea, creating a blockchain-based system of applications that could carry out financial (and other) functions, called Ethereum \citep{buterin_2022}.

Unlike Bitcoin addresses, which store information on so-called Unspent Transaction Outputs, Ethereum addresses store account information like balances as well as code for smart contracts. Smart contracts are computer programs that carry out certain actions upon completion of certain conditions specified within them. There are two types of Ethereum accounts: externally owned accounts (which the owner’s private key controls) and contract accounts (which the smart contract code controls) \citep{buterin_2022}.

\subsubsection{Ethereum Transactions}
Ethereum transactions are essentially cryptographically signed data packages sent from an externally owned account to a recipient, and contain the signature of the sender, the value to be transferred, and a value known as the “gas fee” for the transaction. In Ethereum, users must pay these gas fees to reflect the computational power required to execute the transaction. The fees are paid in Ethereum’s native cryptocurrency, Ether (ETH), which powers the Ethereum ecosystem. This is another difference between Ethereum and Bitcoin—rather than being a store of value like Bitcoin, ETH is “fuel” for the system \citep{buterin_2022}. Figure \ref{fig1} depicts the process of executing an Ethereum transaction \citep{ethereumorg_transactions_2023}. 

At the time of our research Ethereum used proof-of-work (PoW), like Bitcoin, as the consensus mechanism for executing these transactions. Ethereum moved to proof-of-stake (PoS), an alternative consensus mechanism, in September 2022 \citep{noauthor_merge_2022}. In contrast to PoW, wherein validators execute transactions and secure the network by competing to solve computationally hard puzzles, PoS requires would-be validators to lock ETH as collateral; validators who do so are chosen at random to execute transactions and create blocks.

\subsubsection{Ethereum Applications}

Applications are a key part of the Ethereum ecosystem and the primary characteristic differentiating Ethereum from Bitcoin. The Ethereum Virtual Machine (EVM) uses a stack-based bytecode programming language to execute these applications \citep{buterin_2022}. Smart contract code for Ethereum applications is written in a programming language called Solidity and then compiled into the bytecode. The bytecode executes various operational codes (opcodes), which provide computational instructions to the EVM \citep{wood_ethereum_2021, cai_decentralized_2018}.

\begin{figure*}
    \centering
    \includegraphics[width=14.5cm]{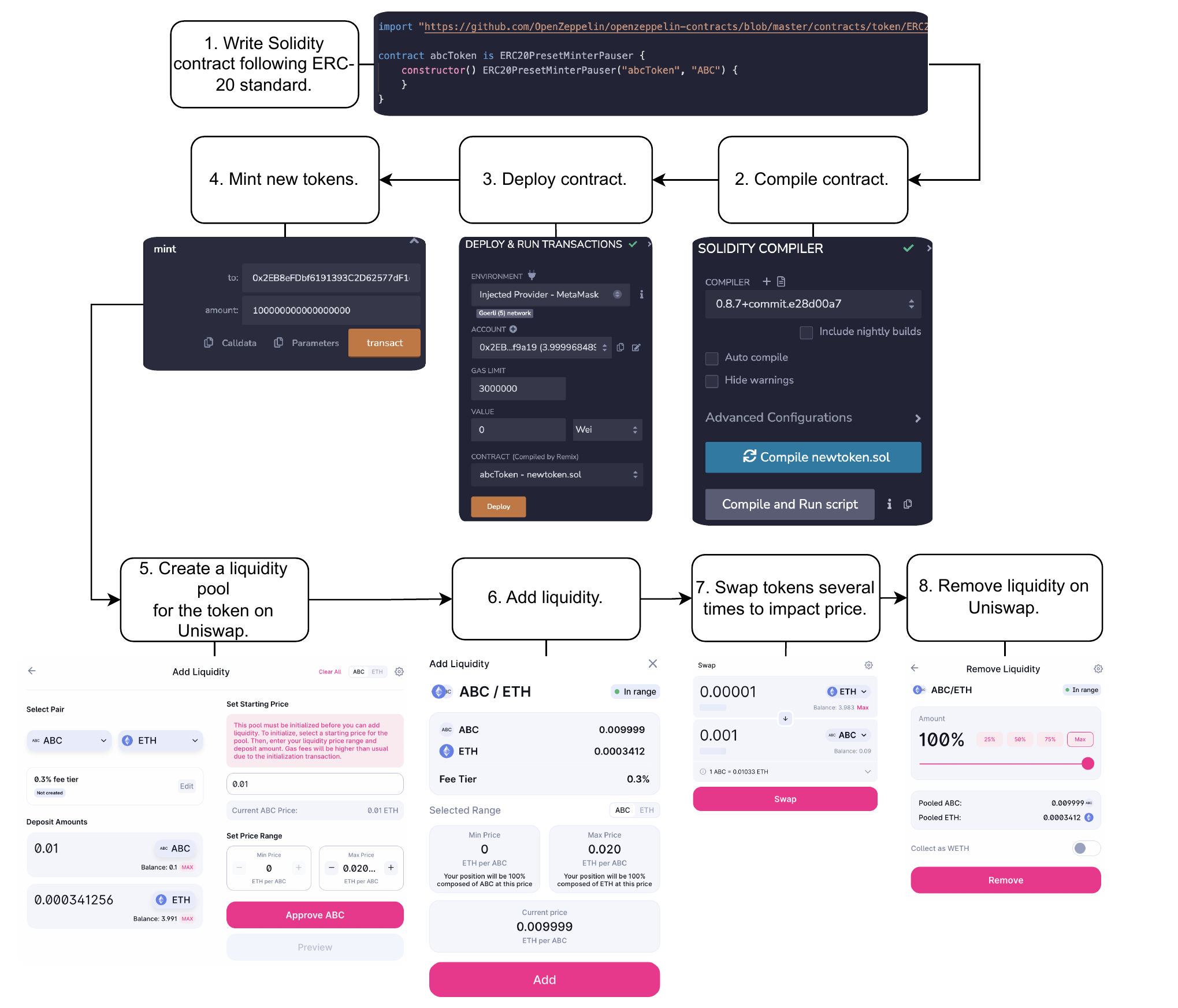}
    \caption{Schematic diagram of the process of creating an ERC-20 token and carrying out a rug pull/pump-and-dump scheme.\\ 1. Write ERC-20 token contract in Solidity, using Open Zeppelin library.\\ 2. Use Solidity compiler to compile code so it can be executed by the EVM.\\ 3. Deploy contract to Ethereum blockchain (in this case, the Goerli test network).\\ 4. Mint new tokens and send to a specific address. \\5. Initiate trading for new token on Uniswap.\\ 6. Add liquidity to enable trading between ETH and ABC token.\\ 7. Swap newly minted ABC tokens for ETH to “pump the price" and then back again (for a profit).\\ 8. Remove liquidity from Uniswap to halt trading of ABC token.}
    \label{fig2}
\end{figure*}

Ethereum has three primary types of applications: financial applications, semi-financial applications, and non-financial applications \citep{buterin_2022}. In this paper, we focus on the financial applications. “Decentralized Finance,” or “DeFi” for short is one category of financial applications built on Ethereum (though, of course, DeFi also exists on other blockchains). DeFi is a system of smart contract-enabled financial products and services like currency exchange, loans, and derivatives, which are built and delivered in an open-source, permissionless, and decentralized way with smart contracts. At all times, users retain custody of their own funds \citep{schar_decentralized_2021}. For a full introduction to Ethereum-based DeFi and its current, primary product offerings, see \citep{https://doi.org/10.48550/arxiv.2112.02731}. 

Tokens are a key part of the Ethereum ecosystem. These include “sub-currencies” and utility tokens \citep{buterin_2022}. Colloquially and collectively, these are called “altcoins”. Many Ethereum-based DeFi projects have associated governance or utility tokens which follow the ERC-20 standard. The ERC-20 token standard specifies various characteristics which developers must define for tokens to ensure their interoperability with the Ethereum ecosystem. Governance tokens (i.e., the UNI token for the Uniswap decentralized exchange) allow participants to vote on the future of projects and project treasury allocation. The process for creating ERC-20 tokens is shown in Figure \ref{fig2} in steps 1-4 \citep{bachini_how_2021}. For full details on Ethereum and the ERC-20 standard, see \citep{wood_ethereum_2021} and \citep{pomerantz_ori_erc-20_2021}, respectively. 

\begin{figure*}
    \centering
    \includegraphics[width=13cm]{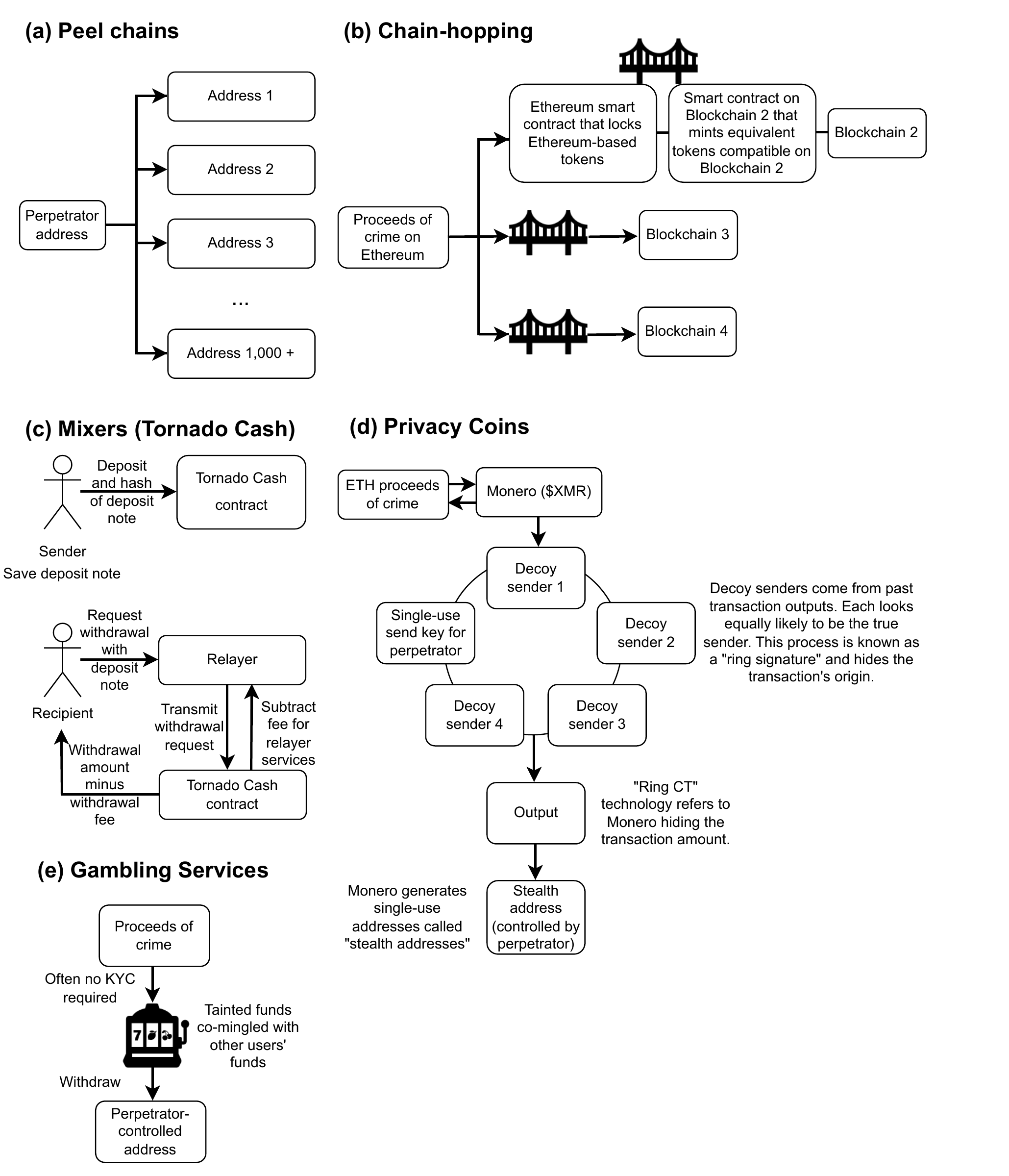}
    \caption{Cryptocurrency-based money laundering techniques: \\
    (a) Peel chains, meaning creating various addresses to which to transfer small amounts of cryptocurrency quickly; \\
    (b) Mixers, like Tornado Cash; \\
    (c) Chain-hopping, or moving cryptocurrrencies among various blockchains using blockchain bridges; \\
    (d) Privacy coins, such as Monero; and \\
    (e) Gambling services.}
    \label{fig3}
\end{figure*}

\subsection{DeFi fraud and money laundering}

Empirical research has chronicled various types of fraud across DeFi, including market manipulation \citep{hamrick_analyzing_2021, mazorra_not_2022, qin_attacking_2021, victor_detecting_2021, wang_towards_2021}, fraudulent investment schemes \citep{xia_demystifying_2021, mazorra_not_2022}, and exit scams (called “rug pulls") \citep{xia_demystifying_2021, mazorra_not_2022}. \citet{xia_demystifying_2021} describe a typical rug pull scam. A scammer creates a token and provides liquidity on Uniswap to trade this token with a popular cryptocurrency. They use social media and advertisements, often on Telegram, to find victims. Then, the scammer removes all tokens from the liquidity pool, leaving the victims holding the now-defunct token. They note that rug pulls are often combined with pump-and-dump schemes, whereby scammers manipulate the price of their token before they sell (which then crashes the price).

Rug pulls are one of the most costly types of securities fraud across DeFi overall, with victims losing \$2.8 billion (of the total \$7.8 billion lost to DeFi scams) in 2021 to rug pulls \citep{chainalysis_2022_2022}. Specific aspects of the DeFi ecosystem facilitate these frauds like price oracles \citep{gudgeon_decentralized_2020, schar_decentralized_2021} and flash loans \citep{caldarelli_blockchain_2021, gudgeon_decentralized_2020, qin_attacking_2021, wang_towards_2021, xu_sok_2022}. Further definitions of these types of fraud can be found in \citep{kamps_cryptocurrency_2022}.

There is sparse peer-reviewed research on the use of DeFi specifically in money laundering, though private companies like Chainalysis have reported on its use \citep{chainalysis_2022_2022}. Their 2022 Crypto Crime Report estimates that addresses they have tagged as “illicit" sent \$900 million to DeFi protocols in 2021. Furthermore, they allege that North Korean hackers are using DeFi on various blockchains and mixers to launder the proceeds of their DeFi hacks and highlight an example of an unspecified attacker using blockchain bridges and mixers like Tornado Cash to launder the proceeds of another hack \citep{chainalysis_2022_2022}. Blockchain bridges allow users to move cryptocurrencies from one blockchain to another, for example, from the Ethereum blockchain to the Polygon blockchain \citep{mccorry_sok_2021}. Most commonly, bridges utilize smart contracts---a user sends the tokens they wish to “bridge" to the smart contract on the originating blockchain. These tokens are locked in the smart contract on the originating blockchain; a smart contract on the destination blockchain then mints equivalent tokens on the destination blockchain, which the user can then use on that blockchain \citep{belchior_survey_2021}. A visual representation of this process can be found in Figure \ref{fig3}(b). 

Mixers are a type of privacy-preserving technology and have been used launder proceeds of crime \citep{akartuna_preventing_2022}. Tornado Cash---one of the most popular Ethereum smart contract mixers---is the most relevant for our purposes \citep{beres}. Users send funds to the Tornado Cash smart contract and, in turn, generate a cryptographic note. When they want to withdraw their funds, they use this deposit note and zero knowledge proofs (which allow one to prove their knowledge of something without revealing the thing itself) to prove the deposit is theirs \citep{chainalysis_team_understanding_2022, wade_how_2022}.

A relayer service further ensures anonymity. Relayers are a decentralized network of users who manage mixer withdrawals from the Tornado Cash smart contract---they pay the gas fees required to conduct the withdrawal transactions (and also deduct a fee for themselves from the withdrawal itself). This inhibits linkages being made between the deposit and withdrawal accounts because the recipient is not the one paying the withdrawal gas fee \citep{chainalysis_team_understanding_2022}. A visualization of the Tornado Cash mixing process is in Figure \ref{fig3}(c). 

While these figures and case studies are a useful starting point for understanding the DeFi-based money laundering more generally, we note that Chainalysis research is (a) not peer-reviewed, and (b) published primarily for marketing purposes. 

Despite the absence of DeFi-specific money laundering research, academic work has long discussed the use of cryptocurrencies more broadly for money laundering. In general, cryptocurrency money laundering fits into the traditional money laundering stages of placement, layering, and integration \citep{desmond_evaluating_2019}; however, the placement process is only relevant in cases where a criminal is seeking to launder proceeds of non-cryptocurrency-native offenses (as, otherwise, the funds are already present in the cryptocurrency ecosystem).  The layering process---where criminals attempt to hide the path their cryptocurrencies take \citep{albrecht_use_2019}---employs various devices including:

\begin{itemize}
    \item Peel chains, meaning creating various addresses to which the criminal transfers smaller amounts of cryptocurrencies \citep{tsuchiya_how_2021, pelker_using_2021}.
    \item Mixers \citep{akartuna_preventing_2022, natarajan_cryptocurrencies_2019}, such as Tornado Cash, as discussed above.
    \item Exchanging cryptocurrencies for other cryptocurrencies and moving existing cryptocurrencies to other blockchains (“chain-hopping"), generally through numerous, quickly-executed transactions \citep{raza_study_2021, pelker_using_2021, natarajan_cryptocurrencies_2019}. Chain-hopping requires the use of blockchain bridges, as described above.
    \item Privacy coins and blockchains \citep{raza_study_2021, natarajan_cryptocurrencies_2019, akartuna_preventing_2022}, such as Monero. Monero utilizes several measures to enhance the anonymity of their users and their transactions. It uses “ring signatures" to hide transactions' origins, which involve combining decoy transaction outputs from previous transactions. Each of the decoy signatures, along with the single-use send key generated for the transaction, look equally likely to an outside observer to be the true sender. Monero also employs “Ring CT" technology which hides transaction amounts and single-use addresses called “stealth addresses" \citep{monero_about_nodate, monero_ring_nodate-1, monero_ring_nodate}. 
    \item Gambling services \citep{fanusie_bitcoin_2018}, which co-mingle tainted funds with other customers' funds.
\end{itemize}
We depict these money laundering techniques visually in Figure \ref{fig3}.

The criminal completes the integration process, which entails using the funds for non-nefarious purposes and co-mingling them with other funds which are not proceeds of crime \citep{albrecht_use_2019, natarajan_cryptocurrencies_2019}. This could involve transferring the proceeds of crime to government-issued fiat currencies or conducting further cryptocurrency investment activities \citep{natarajan_cryptocurrencies_2019}.

\subsection{Detecting DeFi Fraud}

There is limited literature devoted to using computational methods to detect fraud or other illicit activity in DeFi on Ethereum specifically. This research uses various machine learning algorithms to detect smart contract Ponzi schemes on Ethereum including using long short-term memory neural networks \citep{wang_ponzi_2021, hu_transaction-based_2021} and an “anti-leakage" ordered boosting model \citep{fan_-spsd_2021}. Two other studies \citep{xia_demystifying_2021, mazorra_not_2022} also use machine learning models to detect scam tokens on the Uniswap decentralized exchange.

\subsubsection{Rug Pulls}

As discussed above, rug pulls are a type of exit scam. The perpetrator creates a token and then adds liquidity on a decentralized exchange to enable users to trade this new token with another, existing cryptocurrency (most commonly a reputable cryptocurrency like Ethereum) \citep{xia_demystifying_2021}. Uniswap is one of the most popular decentralized exchanges and anyone can add token trading pairs to it (and, unfortunately, many that are added end up being rug pulls) \citep{coingecko_top_2023}. The scammer then recruits victims, often on social media or messaging apps like Telegram, convincing them to buy the token. Scammers employ various tactics at this point (described below), but, ultimately, the result of their actions is that victims are left holding a worthless token they are unable to trade \citep{xia_demystifying_2021}. See Figure \ref{fig2} for more details on the rug pull execution process. 

\citet{xia_demystifying_2021} find more than 10,920 rug pull scams on the Uniswap decentralized exchange (about 50\% of the listed tokens at the time) with profits of at least \$16 million (though they provide limited detail as to how they calculate this profit). They highlight the prevalence of so-called “collusion addresses” in coordination with scam creators and the existence of token smart contract back-doors which further perpetrators' profits. The study identifies 39,762 “potential victims” of these scams. 

Their ground truth comes from manually selected phishing tokens and tokens labelled as scams on the Ethereum blockchain explorer Etherscan. The authors subsequently use guilt-by-association heuristics to expand their data set. They build classifiers (a random forest model performed best) including temporal, transaction, investor, and Uniswap-based features.

\citet{xia_demystifying_2021} describe a typical rug pull scam, citing the RADIX token as an example of this tactic. However, the authors are not systematic in conducting the analyses which led to this conclusion (beyond the use of their machine learning classifier).

\citet{mazorra_not_2022} build on \citeauthor{xia_demystifying_2021}'s (\citeyear{xia_demystifying_2021}) work, adding 16,037 more tokens to their Uniswap scam token data set. They also develop machine learning models with smart contract and investor distribution features, which they assert allows them to preemptively detect rug pulls. They further advance \citeauthor{xia_demystifying_2021}'s work by systematizing profit calculations for these schemes. 

\begin{figure*}
    \centering
    \includegraphics[width=8.5cm]{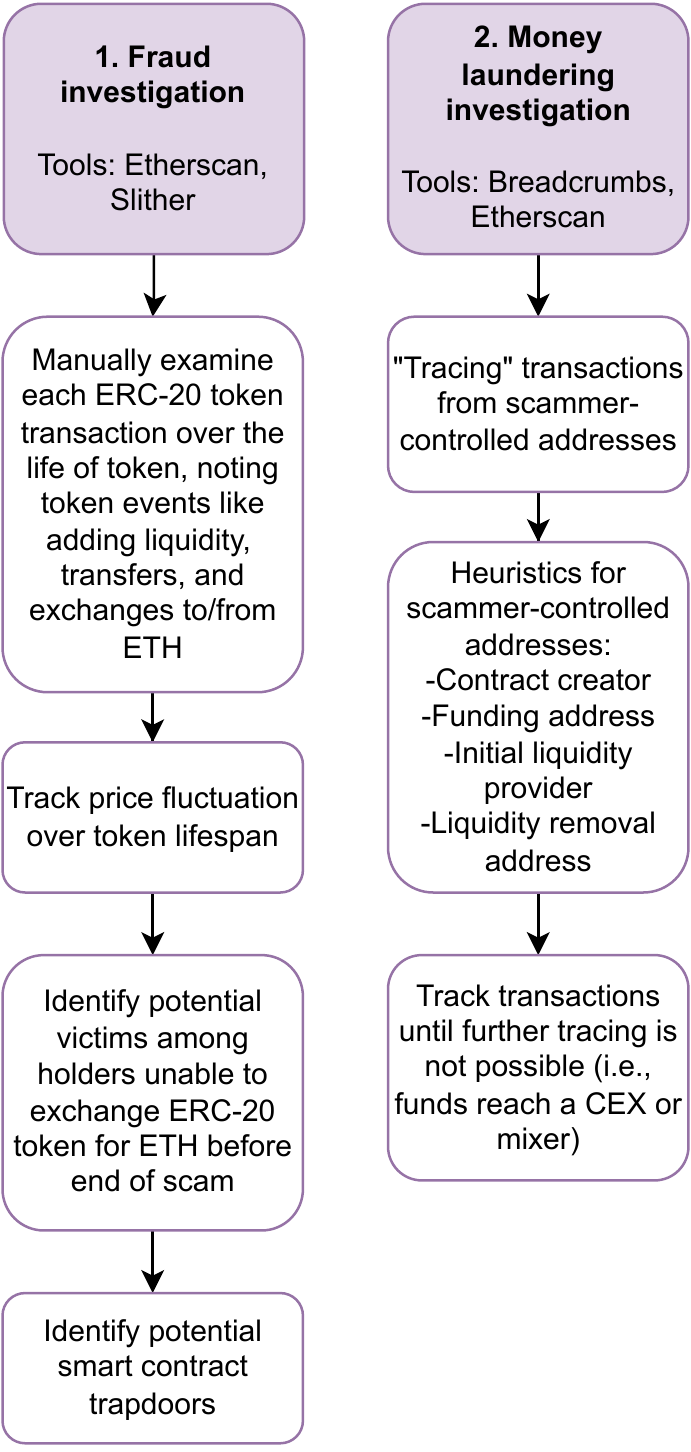}
    \caption{Investigative methods and open-source investigative tools.}
    \label{fig4}
\end{figure*}

\citet{mazorra_not_2022} claim to manually analyze the data from their classifier to develop typologies of rug pull scams on Uniswap. However, they do not offer details on how they conduct this analysis, nor do they indicate that it was conducted systematically. They identify three rug pull typologies:

\begin{itemize}
    \item \textbf{Simple rug pulls,} in which the developer simply removes liquidity from Uniswap \citep{mazorra_not_2022} (akin to a “fast rug pull" as identified by \citet{mackenzie_criminology_2022}).
    \item \textbf{Sell rug pulls,} whereby a scammer creates a token and adds a portion of liquidity to the Uniswap protocol. Victims begin participating in the scam, swapping their legitimate tokens for the scam one. At some point, the fraudster swaps the remaining supply of the token for the legitimate token paired with it in the Uniswap liquidity pool. In some cases, the scammer can also recover their original liquidity too \citep{mazorra_not_2022}. This type of rug pull is slightly harder to identify and calculate profits of than simple ones. Mackenzie \citep{mackenzie_criminology_2022} refers to this as a “slow rug pull" and highlights the psychological manipulation tactics scammers may use to further their scam, such as reassuring investors on Telegram or Discord and encouraging them to purchase more tokens at the now lower price. 
    \item \textbf{Smart contract trapdoor rug pulls,} which embed attack vectors in the token smart contract code. These are the most difficult to identify and prevent \citep{mazorra_not_2022}. There are several such functions that can be coded into smart contracts, such as automatically charging investors to swap their tokens (advance fee tokens) and prohibiting holders from selling the tokens \citep{xia_demystifying_2021}. \citet{mazorra_not_2022} use a tool called Slither to identify such issues in smart contract code, which we also utilize in this study.
\end{itemize}

\section{Method}

We conducted detailed, manual investigations of five ERC-20 tokens using open-source investigative tools to extract evidence of fraudulent activity and uncover subsequent money laundering tactics. From on-chain data, we identified patterns in DeFi fraud and money laundering offending. We show our full investigative process in Figure \ref{fig4}.

We chose to investigate five tokens based on the level of granularity with which we planned to conduct our investigations. For each token, we manually inspected at least hundreds (and, in some cases, thousands) of individual transactions and the components thereof, as well as the addresses that conducted them (alongside their transactions, which were, again, generally quite numerous) to trace the scheme and associated money laundering. For reference, a full-scale securities fraud investigation, carried out by a team of investigators, tends to take several months or even years \citep{us_securities_and_exchange_commission_investor_2014}.

\subsection{On-chain investigations}
Following \citep{dyson_scenario-based_2020}, we began our investigation with Etherscan. Etherscan has both a web-based version and a publicly-available API. For each contract, Etherscan displays various information from the Ethereum blockchain, including the contract creator, contract balance, ERC-20 token transactions, and contract events, among other information. It also provides analytic information, for example, regarding the contract's highest and lowest balances, and any comments from the community. On the token page (reachable from the contract page), Etherscan shows the price, fully diluted market cap, maximum total supply, transfers, current holders, decentralized exchange trades, and contract source code in Solidity. The page also displays any token reputation tags submitted by the Ethereum community and analytic information on the amount of money in the contract, the number of unique senders and receivers, and the number of token transfers. For further details on the information Etherscan and its API provide and the usefulness of this information for blockchain forensic investigations, see \citep{dyson_scenario-based_2020}. In Appendix A, Figure \ref{fig9} we show the token page and the contract page for the UNI token.

For the purposes of our investigation into potential fraudulent activity, we were concerned with the ERC-20 token transfers. We manually examined each transaction, noting its actions and the addresses involved to develop a picture of the scheme. We conducted our analysis in two parts: (1) investigating the scheme itself, (2) tracking the money laundering process.

\subsubsection{Fraud investigation}

In the first step of the analysis phase of our investigations, we were primarily concerned with the occurrence of token events---namely, events like adding liquidity, token transfers, exchanges to and from ETH---as well as price fluctuations as these events took place. We examined each transaction involving the token in question in detail.

At this stage, we also identified potential victims of the scam based on which token holders were unable to exchange their ERC-20 tokens for ETH or another reputable cryptocurrency before the end of the scam. We note that addresses that generally held many non-reputable ERC-20 tokens could, in fact, have created various other scam tokens (\citep{xia_demystifying_2021} found that 24\% of scammer addresses were repeat offenders) or may not, in fact, be victims at all, but rather active participants seeking high return in exchange for participating in a high-risk investment. These types of traders are called “degens" in the cryptocurrency community, which is short for the phrase “Decentralized Finance Degenerates" \citep{nabben_web3_2023}.  Finally, when the perpetrator of a scam removes liquidity for an ERC-20 token on a decentralized exchange, they receive both the remaining ERC-20 token and the token with which it is paired (usually ETH). Therefore, while they are also “stuck" holding the worthless ERC-20 token, they are, of course, not victims.

Following \citep{mazorra_not_2022}, we used Slither \citep{feist_slither_2019} to identify potential smart contract trapdoors among the tokens we analyzed. Slither is “a Solidity static analysis framework”. Since the original paper detailing Slither was published, the package now runs 80 different detectors, including vulnerability, informational, and optimization detectors. This includes vulnerabilities including re-entrancy vulnerabilities and contract name reuse \citep{crytic_slither_2022}.

\subsubsection{Money laundering investigation}

The final step of our analysis involved “following the money" to identify where funds exchanged for ETH from the tokens analyzed in our fraud investigation ended up and the path they travelled, a process known as “tracing" \citep{pelker_using_2021, dyson_scenario-based_2020}. This required us to use various heuristics to identify addresses likely to be associated with the scammer. We assumed the contract creator (and any wallets that funded the address that created the contract) were associated with the fraudster because only the perpetrator or someone colluding with them could have created the fraudulent token. The address which provided the initial liquidity for the token to a decentralized exchange and to whom the majority of the liquidity was ultimately removed from said exchange (if applicable) are scammer-controlled for the same reason. Finally, in some cases, addresses that managed to exchange the scam token for ETH at the token's highest value could be associated with the scammer, though they could also simply be lucky participants in the scam (because the coordinator of the scheme to “pump" the price of their token would be the only party able to perfectly time the highest value of the token \citep{kamps_moon_2018}. Furthermore, these addresses may show a spike in their value at the time of or immediately following the scam; unless a scam was particularly poorly executed, it is likely the perpetrators themselves would extract the most profit from it. We focused our attention on addresses which received the highest value of funds from the scheme for this reason.

We note that \citet{xia_demystifying_2021} describe similar heuristics for what they term “collusion addresses", including those who add initial liquidity on Uniswap, those to whom liquidity was removed on Uniswap, those who exchange tokens for the scam tokens, and those who exchange the scam tokens for legitimate ones. However, we note that only those addresses falling into the first two categories are undoubtedly scammers, which is why we provide further specificity in the heuristics detailed above. While those who are simply exchanging tokens may be engaging in wash trading (which \citet{victor_detecting_2021} suspect may be an issue on decentralized exchanges like Uniswap), we are unable to verify this. 

For the money laundering portion of our analysis, we also utilized Breadcrumbs, a blockchain visualization tool.\footnote{\footnotesize{https://www.breadcrumbs.app/}} Breadcrumbs' Investigation Tool is an open tool that generates visual representations of the flow of funds to and from cryptocurrency addresses. We note that using an openly available tool like Breadcrumbs allays \citet{pelker_using_2021}'s concerns about the potential (though not “insurmountable") litigation risks of using certain popular subscription-based blockchain analytics tools that “incorporate sensitive or proprietary techniques that cannot be readily presented in open court".
Breadcrumbs shows the originating and destination addresses for funds, amounts sent, balances, and other information for each address. For very active addresses, we focused our attention on shorter periods immediately after the scam period, when scammers would be most likely to move the proceeds of their crimes. Finally, we would expect criminal addresses to cease activity after they laundered their funds; therefore, addresses that are still active are less likely to be associated with the scammers. However, those that are inactive are not necessarily scammers; they may just not be participating in trading due to market conditions or for other reasons. 

Using the Breadcrumbs tool, we followed the flow of funds across various addresses until they reached either (a) an address tagged as a centralized exchange, or (b) a mixer like Tornado Cash. Once funds reach these destinations, we are unable to trace them further (though, in the case of centralized exchanges, law enforcement intervention could elicit further information, as many centralized exchange services require customers to submit Know Your Customer information upon registration)\citep{dyson_scenario-based_2020}.\footnote{\footnotesize{We also attempted to use K-Means clustering on the addresses involved in the schemes and any addresses with which they interacted to determine whether any of the addresses may be controlled by the same person. However, ultimately, this clustering split the addresses into two classes, (1) those who participated in the scheme, and (2) those that did not. We note that \citet{mazorra_not_2022} found as well that the addresses involved in the rug pulls they examined also evaded existing Ethereum address clustering techniques.}} 

We note that Etherscan also indicates where wallet addresses are also found on blockchain explorers for other blockchains. This could even be the case for the tokens themselves. However, for the purpose of this study, we only examine activity on the Ethereum blockchain. It would be useful for future research to explore automated detection and, subsequent, manual investigation of DeFi tokens on other blockchains, particularly given that so-called “chain-hopping" is a well-known cryptocurrency money laundering method \citep{pelker_using_2021}.

\begin{table*}
\caption{Scam token characteristics}
\footnotesize
\label{table1}
\begin{tabular}{|p{2.5 cm}|p{2 cm}|p{2 cm}|p{2 cm}|p{2 cm}|p{2 cm}| p{2 cm}|} \hline
     & Token 1 & Token 2 & Token 3 & Token 4 & Token 5 \\ \hline
    Active period (UTC) & Apr-17-22, 21:07–21:47 & Jun-17-22, 9:08–20:56; final transfer on Jun-20-2022, 12:17 & Apr-13-22, 11:56–12:13 & May-30-22, 7:26–13:36; moved funds May-30-22, 18:38, then Jun-06-22, 11:25 & May-05-22, 13:54–May-09-22, 17:59; two more sell orders May-22-22, 7:04; moved funds Jun-05-22, 10:56 \\ \hline
    Number of transfers & 154 & 94 & 92 & 132 & 500 \\ \hline
    Number of unique addresses (during scam) & 94 & 22 & 41 & 56 & 82 \\ \hline
    Number of remaining holders post-scam (excluding smart contracts) & 77 & 5 & 34\footnote{\footnotesize Includes one null address.} & 35 & 36\footnote{\footnotesize Includes one null address.}\\ \hline
    Total revenue earned swapping token to ETH & 10.57 (\$32,364.07) & 4.91 (\$5,243.83) & 0.21 (\$636.27) & 15.59 (\$28247.37) & 7.11 (\$20,905.04) \\ \hline
    Difference in liquidity between original liquidity provided and liquidity removed & 5.39 (\$16,503.53) & 0.14 (\$149.52)\footnote{\footnotesize Based on final trade and largest amount; did not use remove liquidity function.} & 2.517 (\$7,626.21) & 0.26 (\$471.09) & 0.26 (\$764.46) \\ \hline
    Total ETH spent on token & 15.96 (\$48,867.60) & 2.91 (\$3,107.85) & 2.35 (\$7,120.22) & 15.79 (\$28,609.74) & 8.13 (\$23,904.07) \\ \hline
    Maximum price of token during scam & 4.20e-11 (\$0.0000001 ) & 6.87e-09 (\$0.00001) & 1.45e-08 (\$0.00004 ) & 1.23e-11 (\$0.00000002) & 2.39e-05 (\$0.07) \\ \hline
    Minimum potential profit & 0 (\$0) & 2.14 (\$2,285.50) & 0.38 (\$1,151.35) & 0.1 (\$181.19) & 1.28 (\$3,763.49) \\ \hline
    Maximum potential profit & 15.96 (\$48,867.60) & 5.05 (\$5,393.35) & 2.73 (\$8,271.57) & 15.85 (\$28,718.46) & 7.37 (\$21,669.50) \\ \hline
\end{tabular}
\footnotesize \textsuperscript{5}Includes one null address.
\\
\footnotesize \textsuperscript{6}Includes one null address.
\\
\footnotesize \textsuperscript{7}Based on final trade and largest amount; did not use remove liquidity function.
\\
\footnotesize \textit{All values shown in ETH (USD). USD values given at opening exchange rate from ETH on first day of scam \citep{yahoo_finance_ethereum_2023}.}
\end{table*}
\label{tab1}

\section{Results}

\subsection{Schemes}

All of the five schemes we analyzed were rug pull scams that exhibited pump-and-dump behavior. All the scams had an “unknown" reputation according to Etherscan, indicating that these are as yet unreported. The general pattern of behavior is as follows (and is also depicted in Figure \ref{fig2}):
\begin{enumerate}
    \item Scammer creates set number of tokens.
    \item Scammer enables trading of the new token on Uniswap, creating and funding a liquidity pool for the token/ETH trading pair.
    \item The scammer (through various addresses they likely control), or others they influence, buy the token on Uniswap using ETH, artificially inflating demand for the token and, therefore, its price.
    \item The scammer (or and other traders who manage to time the pump-and-dump scheme correctly) sell the token for ETH on Uniswap. This buying and selling pattern may occur in rapid succession several times.
    \item The scammer removes liquidity from the Uniswap pool, either by using the “remove liquidity" function and sending the remaining funds to an address they control, or swapping the rest of the remaining scam token in the pool to ETH.
\end{enumerate}

Appendix B gives a practical overview of the investigative process for Token 1, including screenshots of the tools we used at various stages of our investigation. Table \ref{table1} highlights various characteristics of the (anonymized) scams we investigated. In terms of the other characteristics of the scams, it is harder to generalize amongst those investigated besides the overall pattern of behavior. The length of the scam ranged from 40 minutes to four days and the number of transfers of each of the tokens between 92 and 500. The percentage of remaining token holders (of all the unique addresses involved) varied between 23\% and 83\%. 

We cannot calculate the profitability of the scams without knowing all of the addresses associated with the perpetrator; however, we estimate the minimum potential profitability, \textit{p}, in the following equation, where \textit{R} is the revenue earned, in ETH, \textit{S} is the total ETH spent, and \textit{L} is ETH liquidity for the token-ETH pool:

\[p=(R-S) + \Delta L\]

\begin{figure}
\centering
    \includegraphics[width=10cm]{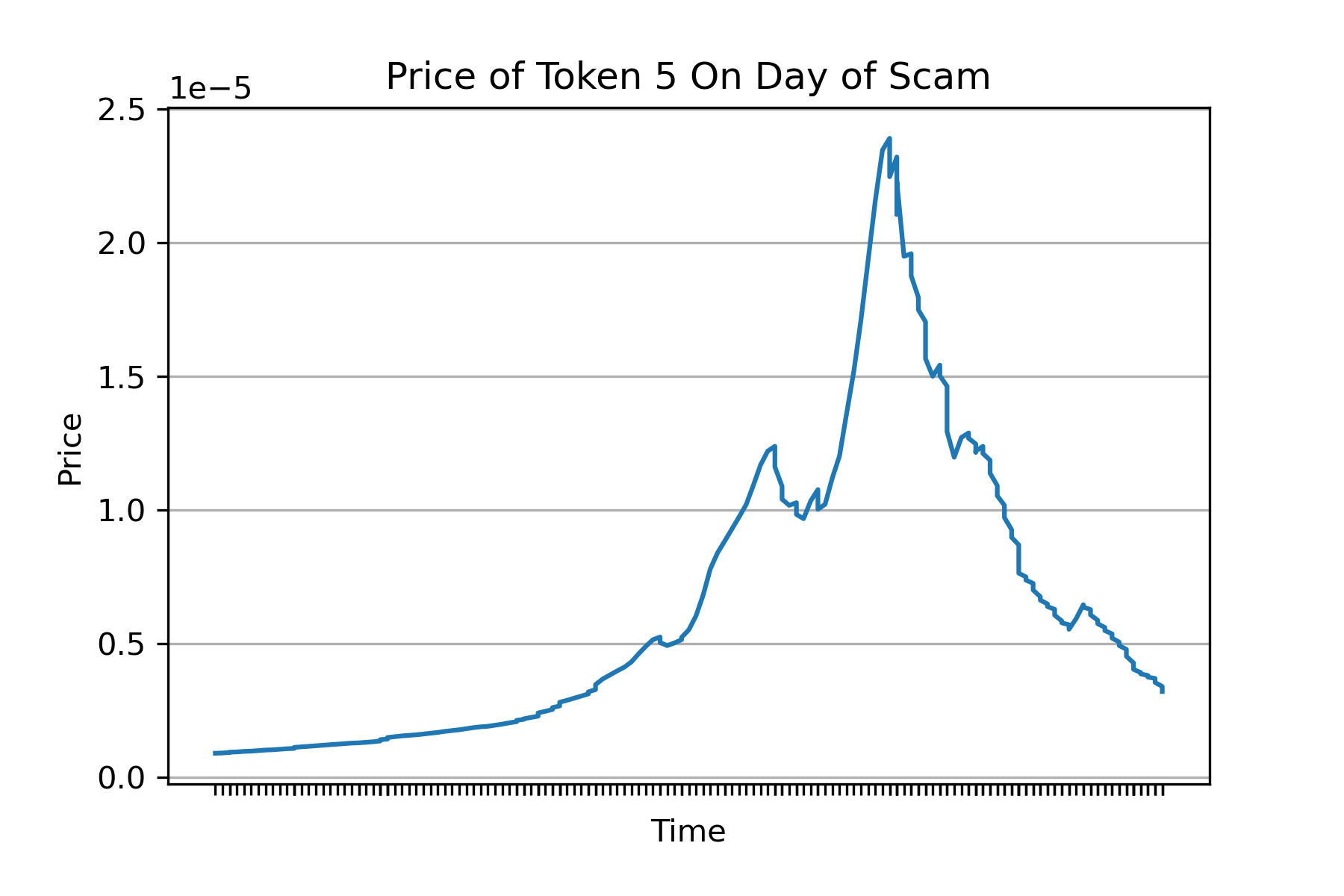}
    \caption{Price of Token 5 throughout fraudulent scheme.}
    \label{fig6}
\end{figure}

\begin{figure}[h!]
    \centering
    \includegraphics[width=10cm]{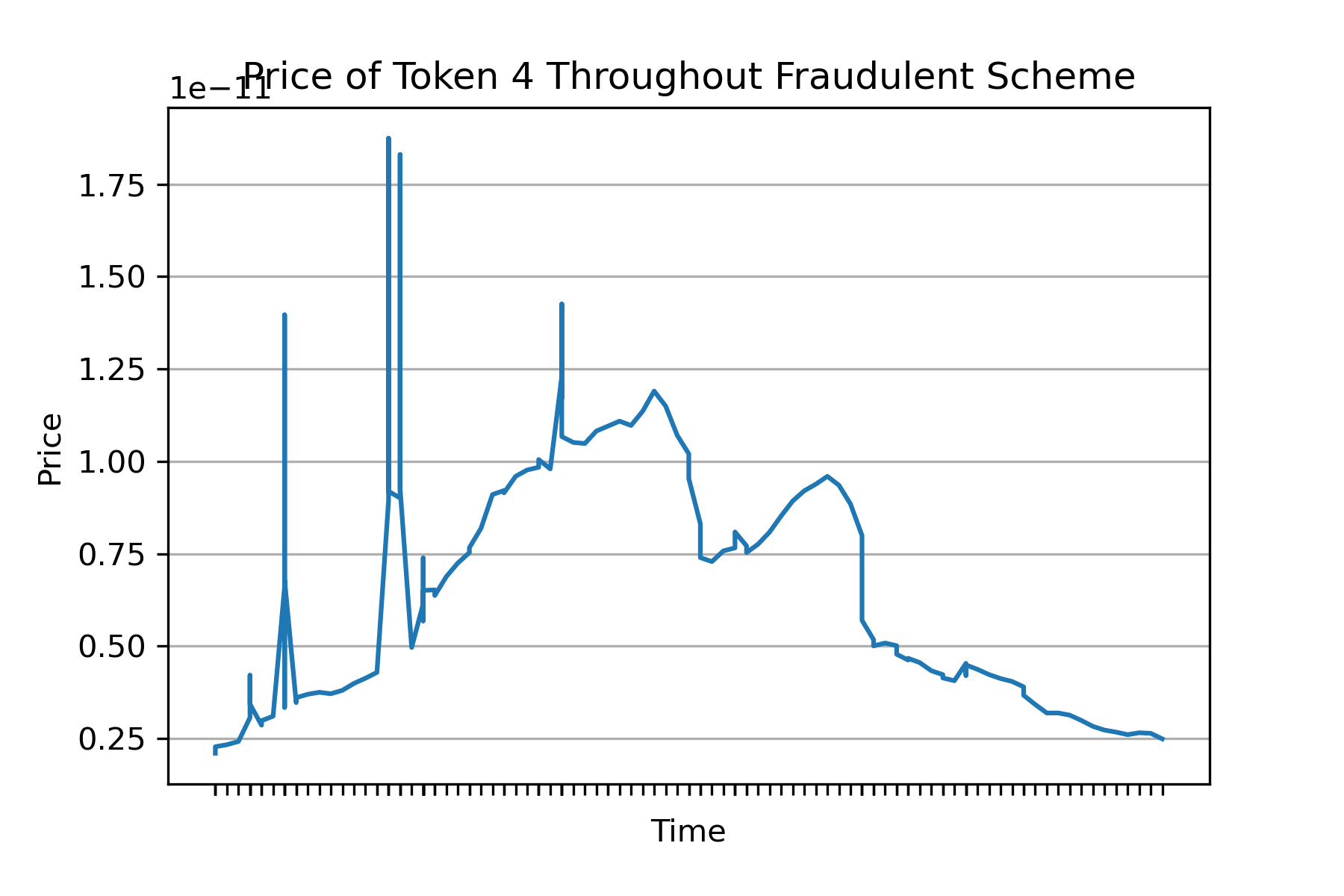}
    \caption{Price of Token 4 throughout fraudulent scheme.}
    \label{fig7}
\end{figure}

We estimate the maximum potential profit, \textit{P}, using the same variables, as:

\[P=R+ \Delta L\]

The ranges for potential profitability vary greatly among the schemes and suggest that some may not have been very profitable at all. Future research could investigate this further.

Figure \ref{fig6} shows the change in the price of Token 5 throughout the scheme. The other tokens' prices showed a similar trend, with the exception of Token 4, which experienced several more peaks and troughs in its price throughout the life of the scam (depicted in Figure \ref{fig7}). This is because there were more sales throughout the life of Token 4 interspersed with the buy orders, rather than a series of several buy orders followed by a series of several sell orders only (as was the case in some other schemes).

\begin{table*}[!h]
\caption{Money laundering schemes}
\footnotesize
\label{table2}
\begin{tabular}{|p{2 cm}|p{1.75 cm}|p{1.75 cm}|p{1.75 cm}|p{1.75 cm}|p{1.75 cm}| p{1.75 cm}|} \hline
     & Token 1 & Token 2 & Token 3 & Token 4 & Token 5 \\ \hline
   Dates scammer wallet active & Apr-17-22--Apr-18-22 & Jun-17-22--Jun-20-22 & Apr-13-22 & May-28-22--Jun-06-22 & May-4-22--Jun-5-22 \\ \hline
   Number of transactions by scammer wallet & 15 & 12 & 10 & 12 & 41 \\ \hline
   Money laundering strategies & Chain-hopping; potentially gambling platforms for other scams & Peel chains & Sending funds through one other address & None with primary wallet & Peel chains \\ \hline
   Cash-out method & Unknown (but uses centralized exchanges in general) & Bitfinex, OkEx, Crypto.com, Gate.io, Bittrex & Coinbase & Binance & Kucoin \\ \hline
   How wallet was funded & Active wallet with ByBit account & Active wallet with ByBit account & Coinbase & Binance & Kucoin \\ \hline
\end{tabular}
\end{table*}
\label{tab2}

\subsubsection{Smart contract analysis}

Of the 24 high-impact vulnerabilities Slither detects, all tokens except Token 4 exhibited only a single vulnerability: re-entrancy vulnerabilities\footnote{\footnotesize{{Re-entrancy attacks exploit a smart contract vulnerability which allows an attacker to call a smart contract multiple times before the contract has finished executing and the state has been updated. An attacker could, for example, call a contract repeatedly to withdraw funds from it several times before the state is updated to reflect the fact that they have already withdrawn their funds.} \citep{crytic_re-entrancy_2018}}}. However, we do not see any of the trapdoor rug pull vulnerabilities cited by \citet{mazorra_not_2022}, such as the TransferFrom vulnerability.

We note that in some of the scams (Tokens 3 and 5) every transfer of the token to ETH seemed to automatically also add liquidity to the Uniswap pool. While this is not inherently malicious (and, likely why Slither does not evaluate these fees), it seems unlikely these fees are advertised in advance. These tokens, therefore, appear to follow the pattern of “advance fee tokens" as described by \citet{xia_demystifying_2021}.

\subsection{Money laundering}
As discussed, we began our money laundering investigations with the addresses which created the scam tokens, added liquidity to Uniswap to trade them, and to which liquidity for the trades was ultimately removed. These are the only addresses we could be certain belonged to the scammer. We also examined how these scammer-controlled addresses (a) were funded, (b) sought to hide the trail of funds earned from the scam, and (c) cashed out to fiat currency after the scam (if applicable).

Table \ref{table2} summarizes the money laundering schemes for each of the tokens analyzed. In all cases, the scammer's wallet was not active for very long (though this varied between a single day and just over a month), and generally did not have many transactions. All of the scammer wallets had some connection to addresses tagged by the community as various centralized exchanges, and received or sent amounts to them that would likely require them to provide KYC information. This is an avenue law enforcement would be able to follow. 

The tactics these addresses used to launder funds ranged. In some schemes (Tokens 3 and 4), no specific laundering techniques appear to have been employed. In the case of Tokens 2 and 5, the scam wallet sent small amounts of ETH to various addresses they seemingly controlled in an attempt to obfuscate the trail of funds. Finally, one scheme (Token 1) used chain-hopping (sending tokens to another blockchain via the Synapse bridge) to hide the trail of their funds. Following the funds on other blockchains was outside of the scope of this study, so we instead examined some of the other addresses with which the scammer interacted in more detail.

\begin{figure}[h!]
    \centering
    \includegraphics[width=11cm]{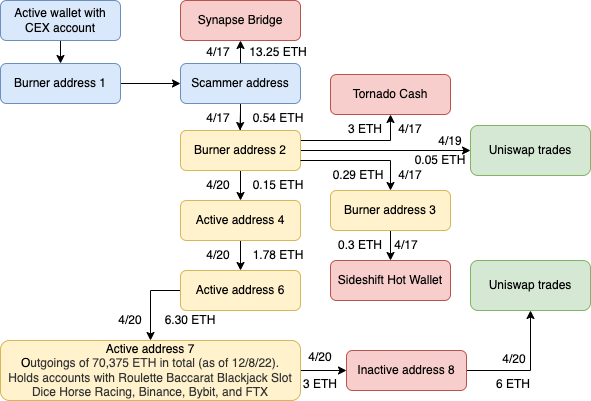}
    \caption{Token 1 money laundering scheme.}
    \label{fig8}
\end{figure}

Figure \ref{fig8} depicts the money laundering activity related to Token 1. The wallet was initially funded by what we refer to as a “burner address", meaning an address created only for a discrete purpose, after which it becomes inactive. Burner addresses may suggest nefarious activity, but could also be used for legitimate purposes (such as privacy protection or for security when interacting with untested dApps or tokens that could have trapdoors in their code). This money was, in turn, funded by an active address that appears to have received money from a ByBit account. In the case of Token 1, while most funds went to this blockchain bridge, some money was sent to another burner address. This address sent funds to another burner address, which traded on Uniswap and also sent money to the mixer Tornado Cash. Still other funds went to another burner address, which, on the day of the Token 1 scam, sent 6.3 ETH to an active address (with 70,375 ETH in outgoings throughout its existence). These funds are unlikely to be the proceeds of the Token 1 scam due to their high value relative to the maximum potential profit from scam 1, but could potentially be from other fraudulent schemes. Some funds were then sent from this address to a gambling platform and two centralized exchanges, in amounts that would legally require them to hold KYC information about the scammer in many jurisdictions. This behavior also suggests the scammer may use gambling platforms to launder other funds. 



In the case of Token 2, various addresses sent funds to one another, including addresses where funds were received and then immediately sent out to another address. The address that funded the address that created the scheme seems to have been used to cash out the proceeds. This wallet is still active and has made more than 100,000 transactions. Its highest balance was 17,804.31 ETH in September 2022. After the scam, the wallet's balance dwindled for several days, before rising again a week later (potentially from proceeds of another scam). This wallet sent large amounts of funds to various centralized exchanges (much more than the likely proceeds of the Token 2 scam), including Bitfinex, OkEx, Crypto.com, Gate.io, Bittrex. Since these transactions are co-mingled, it is unclear to which centralized exchange the proceeds of the Token 2 scam, specifically, went. 

The addresses responsible for creating Tokens 3 and 4 did not participate in sophisticated money laundering activity. In fact, in the case of Token 3, the scammer address was funded by a Coinbase account, before sending funds to another account, which then sent funds to Coinbase. 

In the case of Token 4, the wallet initiating the scam was funded by Binance and then sent funds to Binance a few days after trading of the token ended. Since this coin had more peaks and troughs in its price, it is likely that more addresses were involved, perhaps as part of a coordinated pump-and-dump scheme. However, many of the addresses involved held and traded hundreds of extremely low-value altcoins. This suggests that they are either serial scammers who have conducted similar schemes across many different coins, or that they are merely opportunistic traders. Traders who trade risky altcoins in the cryptocurrency space are generally referred to as “degens", and use various tools or programs to identify tokens with a low market capitalization with the potential for large price gains. They generally expect to lose money on some of these trades, while gaining exceptional returns on others. They are aware they are gambling. This was also the case for other addresses that exchanged the scam tokens for ETH throughout the life of these schemes. 

Various addresses involved in trading Token 4 exhibited what we may label as “suspicious" behaviour, but we are unable to confirm they are associated with the scammer. Future research involving computational clustering could address this. Notably, many of the addresses involved with Token 4 utilised Miner Extractable Value (MEV) bots. This suggests that the traders involved were perhaps more sophisticated than in some of the other schemes. MEV refers to Ethereum miners ordering transactions they see in the \textit{mempool} in a block in a way that captures additional profit for the miner \citep{daian_flash_2019}. This may involve tactics such as front-running, backrunning, or sandwich attacks, which combine the two \citep{xu_sok_2022}. Bots can be coded for this purpose and appear to be utilized in this case. However, \citep{mazorra_not_2022} cite an example of a scam token designed to trick MEV bots.

Finally, from June 5, the creator of Token 5 sent small amounts of ETH to 28 different addresses after the scam (totalling 2.8 ETH, with the highest transfer being for 1.23 ETH). These wallets transferred funds among one another and are generally still active. The address that received the 1.23 ETH, sent 6.14 ETH to Kucoin on the same day. This address is still active and has, at various times, had a very high balance (57,342.74 ETH before the scam).\footnote{\footnotesize Notably, one address that earned 0.07 ETH from trading Token 5 for ETH is tagged on the Breadcrumbs application as being on a Uniswap blocklist. However, the address remains active on Uniswap and appears to participate in many pump-and-dump schemes. It is unclear if this address is controlled by the primary scammer.}

\section{Discussion}

\subsection{Key takeaways}
Our findings with respect to our key research questions can be summarised as follows:
\begin{enumerate}
\item Using open-source investigative tools, including Etherscan and Breadcrumbs, to conduct on-chain investigations, proved fruitful in identifying evidence of several rug pull scams and their perpetrators' money laundering tactics, which could be used in prosecuting these crimes. 
\item These open-source investigative tools also successfully revealed some patterns in how DeFi frauds are committed. Our investigations exclusively found rug pull scams which also utilized pump-and-dump tactics. Overall, the schemes were less sophisticated than we expected. 
\item The open-source investigative tools we used showed funds were laundered in these schemes using rudimentary obfuscation techniques, such as peel chains and chain-hopping. Ultimately, most of the proceeds of the scams arrived at centralized exchanges, where we expect they were withdrawn as fiat currency in amounts under the required limit for submitting Know Your Customer information. 
\end{enumerate}
\subsection{Tools to Detect and Investigate DeFi Fraud}

Notably, our investigations (albeit into a limited number of contracts), only revealed rug pulls of Ethereum-based DeFi tokens, which is perhaps less surprising given estimates that 35.9\% of funds lost to DeFi scams in 2021 were as a result of such schemes \citep{chainalysis_2022_2022}. However, the fact that we found these exclusively may suggest that something about them makes them disproportionately obvious. Rug pulls may also be underreported---many cryptocurrency market participants, in fact, consider being “rug pulled" as a rite of passage. It is also unlikely that the figure provided by Chainalysis includes smaller-scale rug pull schemes like those this paper investigates.

Our manual analysis of the subsequent money laundering activity highlighted Ethereum addresses which participated in the purchase of DeFi tokens which, at first glance, appear to exhibit similar behaviour to those analysed in this study. Future research could analyze patterns of behavior among these addresses, namely, whether they are repeat offenders, or merely so-called “degens" looking to invest in high-risk, high-reward tokens. If they are, in fact, repeat rug pull offenders, the value lost to these scams may be much higher than previously reported. 

While the use of Etherscan and Breadcrumbs certainly proved useful in exposing on-chain evidence of multiple rug pull scams, the investigation process proved time-intensive (several full days of work for each token we investigated). Particularly in the money laundering investigation phase, various addresses of interest executed more than 100,000 transactions throughout their existence. While law enforcement agencies generally have a team of investigators to conduct their investigations, the prevalence of rug pulls means that even these resources are insufficient to capture all offending. Therefore, it may be fruitful for future research to explore ways to automate more of this process, such as automatically applying the heuristics we identified as part of the money laundering investigation phase.

Slither offered rudimentary insight into the content of the smart contract codes in question. Further, manual smart contract analysis was outside of the scope of this study, but is a useful avenue for further research. Furthermore, such analysis could feed into more targeted tools for detecting various types of smart contract trapdoor rug pull schemes. 

\subsection{Legal value of evidence and next steps for investigators}

The data extracted using these open-source investigative tools have evidentiary value because they establish a fact pattern of criminal behavior. With support from an expert witness, this would be useful in prosecuting these frauds. Furthermore, because we use openly available tools rather than proprietary “black box" algorithms to arrive at the relevant conclusions, this evidence is more easily explicable in court.

However, to use the evidence we revealed in a prosecution, law enforcement would need to connect the wallets analyzed with “real-world" identities. Investigators could subpoena centralized exchanges to which tainted funds were sent. Even if funds were sent in small enough amounts to evade KYC requirements (which was not the case in many instances), the scammer may have sent funds to a bank account in their name, or used their real email, or real phone number. Some exchanges also collect IP addresses, “browser fingerprints", and other information about customers \citep{coinbase_data_nodate}. This information could be used to issue further subpoenas, for example, of mobile phone carriers or internet service providers. Ethereum wallets communicate with the Ethereum blockchain through a JSON RPC (remote procedure call) server. This server is often delivered through a “proxy node" from a third-party node service provider like Infura \citep{infura_rpc}. The default RPC endpoint for the most popular non-custodial, hot Ethereum wallet (often used to interact with the DeFi ecosystem), MetaMask, is Infura. Infura collects transaction data and user IP addresses, which they retain for seven days unless the user switches their MetaMask RPC endpoint \citep{kessler_consensys_2022}. While there is the possibility that the scammers could use fake KYC information for their exchange accounts, the overall lack of sophistication of their schemes and money laundering methods makes this seem less likely. 

Investigators would also likely seek information elsewhere, such as from Twitter, Telegram, or Discord accounts; and marketing materials and websites. We note that many of the smart contracts list the tokens' Telegram channel and/or Twitter handle before the start of the code. They could also conduct interviews \citep{us_securities_and_exchange_commission_enforcement_2017} and engage further expert witnesses \citep{pelker_using_2021}. \citet{dyson_scenario-based_2020} also offer methods law enforcement could use to crack users' wallet passwords or uncover their seed phrases, which is likely necessary to recover the proceeds of crime.


\subsection{Ethereum-based DeFi fraud}
As discussed in section 4.3, it was somewhat surprising that all of the scams we investigated involved rug pulls and that they only involved Uniswap. Based on the amount of research on automated detection of Ponzi schemes on Ethereum (see, for example \citep{wang_ponzi_2021, hu_transaction-based_2021, fan_-spsd_2021}), we would have expected to see some (though our sample was very small). Furthermore, our sample came from the most recent set of blocks extracted from Ethereum; it is possible that the type of offending has changed over time (given that many of the aforementioned papers rely on data from 2019 \citep{bartoletti_dissecting_2020}). 

Our research complements findings from \citet{mazorra_not_2022} and \citet{xia_demystifying_2021}. We find that the rug pulls we examined are sell rug pulls based on \citet{mazorra_not_2022}'s categorization and that some also appear to employ smart contract trapdoors in their code. Though we did not quantify this, we also found evidence, as \citet{xia_demystifying_2021} did, that those who participated in these schemes seemed to participate in others. However, our examples did not show repeat scam efforts using the same tokens (unlike \citet{xia_demystifying_2021}'s research). \citet{xia_demystifying_2021} also found that 37\% of scams lasted only one hour or less; this was the case for two of the tokens we analyzed, while the other three were slightly longer.

We were surprised by the relative lack of sophistication of these schemes (particularly Token 3). While we are unable to definitively comment on scammers themselves, our findings suggest that they could be relatively unsophisticated, merely copying a low-effort pattern of offending that worked for others. However, we note that, like \citet{xia_demystifying_2021}, we saw evidence of the use of arbitrage bots in some cases, which might point to more sophisticated perpetrators. They found that 27 of the addresses they identified participated in more than 1,000 Uniswap pools, which they identified as the result of using these bots.

\subsection{Ethereum-based DeFi fraud money laundering}
Similarly to the schemes themselves, the money laundering tactics subsequently applied---if they existed at all---were relatively unsophisticated. Known tactics such as chain-hopping and peel chains are present in some schemes \citep{pelker_using_2021}. Our findings are only somewhat consistent with the narrative that “high-risk" exchanges are often used to launder funds \citep{chainalysis_2022_2022}. While some of the exchanges used could be considered slightly higher risk, others, like Coinbase, are publicly listed in the U.S..

\subsection{Victims}
While we have not conducted a detailed analysis of these schemes' victims, we can make some initial comments. There is some question about whether so-called “degen traders" can be considered victims at all. While violations of securities laws are still illegal, the “victims" very likely understood that they were gambling. 

In terms of how scammers may have recruited victims, we can only hypothesize based on the analysis we conducted. Previous research has reported that many pump-and-dump schemes are coordinated on social media or messaging applications like Telegram \citep{xia_demystifying_2021}. Many DeFi users also use tools such as DEX Screener\footnote{\footnotesize{https://dexscreener.com/}} which shows new trading pairs on various decentralized exchanges or automated trading services that often trade these sorts of assets.\footnote{\footnotesize{See, for example, https://3commas.io/}} 

In our manual analysis of the subsequent money laundering activity associated with the scam tokens studied, we noticed Ethereum addresses purchasing other DeFi tokens which, at first glance, appear to exhibit similar behavior to those scam tokens we analyzed in this study. Future research could analyze patterns of behavior among these addresses, namely, whether they are repeat offenders, repeat victims, or merely so-called “degens" looking to invest in a high-risk, high-reward token.

\subsection{Limitations and future research}

The primary limitation of our research was that we could not investigate more individual tokens because the process was so time consuming. However, even with a limited sample, firm patterns emerged. Future research could explore how to automate more of this process and also conduct similar research on other blockchains. Using automated extraction methods on a larger set of tokens could uncover more robust typologies of Ethereum-based DeFi scams. Furthermore, while we attempted to be as systematic as possible in our on-chain analysis, there are still subjective elements of the process, particularly in our investigation of the money laundering schemes (a point \citep{dyson_scenario-based_2020} echoes). Future research could employ various annotators to conduct analysis on the same tokens. Finally, we only used open tools in our analysis. There are other, potentially more powerful, proprietary blockchain analytics tools offered by private companies.

\section{Conclusions}
Fraud across DeFi is a widely-discussed issue. This paper provided various insights about the nature of Ethereum-based DeFi crime and demonstrated how open-source investigative tools can be used to extract evidence of scams on Ethereum which could be used in prosecuting the same. We conducted these investigations in a systematic manner which would be of use to law enforcement and other researchers. Our investigations using these tools revealed evidence of a series of rug pull scams which employed pump-and-dump tactics. We also systematically investigated money laundering tactics following Ethereum-based DeFi frauds. Like the schemes themselves, the money laundering tactics were rather unsophisticated and easily detectable strategies (like peel chains and chain-hopping); in some cases, scammers did very little to hide their crimes. The proceeds of the rug pulls primarily arrived at centralized exchanges, which represents a useful “choke point" for law enforcement to identify DeFi users. Our findings suggest that rug pulls may be a highly detectable and identifiable type of Ethereum-based DeFi scam and that several, smaller-scale rug pulls may be taking place which are not included in mainstream statistics on DeFi-based offending. Further automation of the investigative process proposed in this paper could allow more, even smaller-scale offenders to be prosecuted. 

\pagebreak

\section*{Funding}
This project was funded by UK EPSRC Grant EP/S022503/1 which supports the Centre for Doctoral Training in Cybersecurity at UCL.

\section*{Authors' Contributions}
\begin{itemize}
\item \textbf{Arianna Trozze:} Conceptualization, Investigation, Formal Analysis, Writing – Original Draft, Visualization
\item \textbf{Bennett Kleinberg:} Conceptualization, Writing – Review \& Editing, Supervision
\item \textbf{Toby Davies:} Conceptualization, Writing – Review \& Editing, Supervision
\end{itemize}



\pagebreak
\appendix
\section{Etherscan Token Page and Contract Page}

\begin{figure*}[h!]
    \centering
    \includegraphics[width=10.5cm]{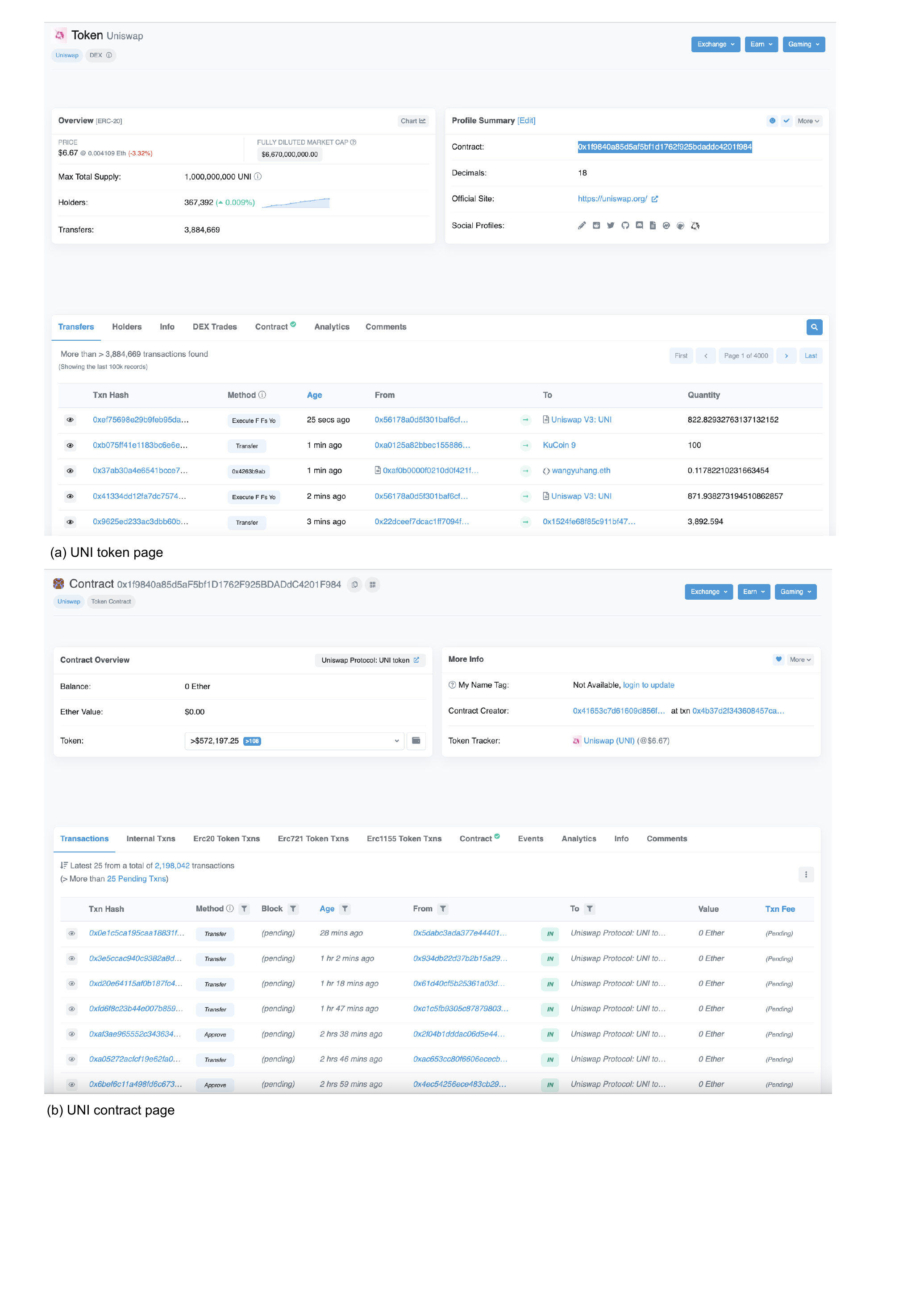}
    \caption{UNI token page (a) and contract page (b) from Etherscan.}
    \label{fig9}
\end{figure*}

\pagebreak

\section{Use of Investigative Tools in Practice}

Below, we show the use of the open-source investigative tools discussed above for selected aspects of our investigation of Token 1. Some information has been redacted in compliance with the requirements of our university's ethics committee. 

\begin{figure*}[h!]
    \centering
    \includegraphics[width=14cm]{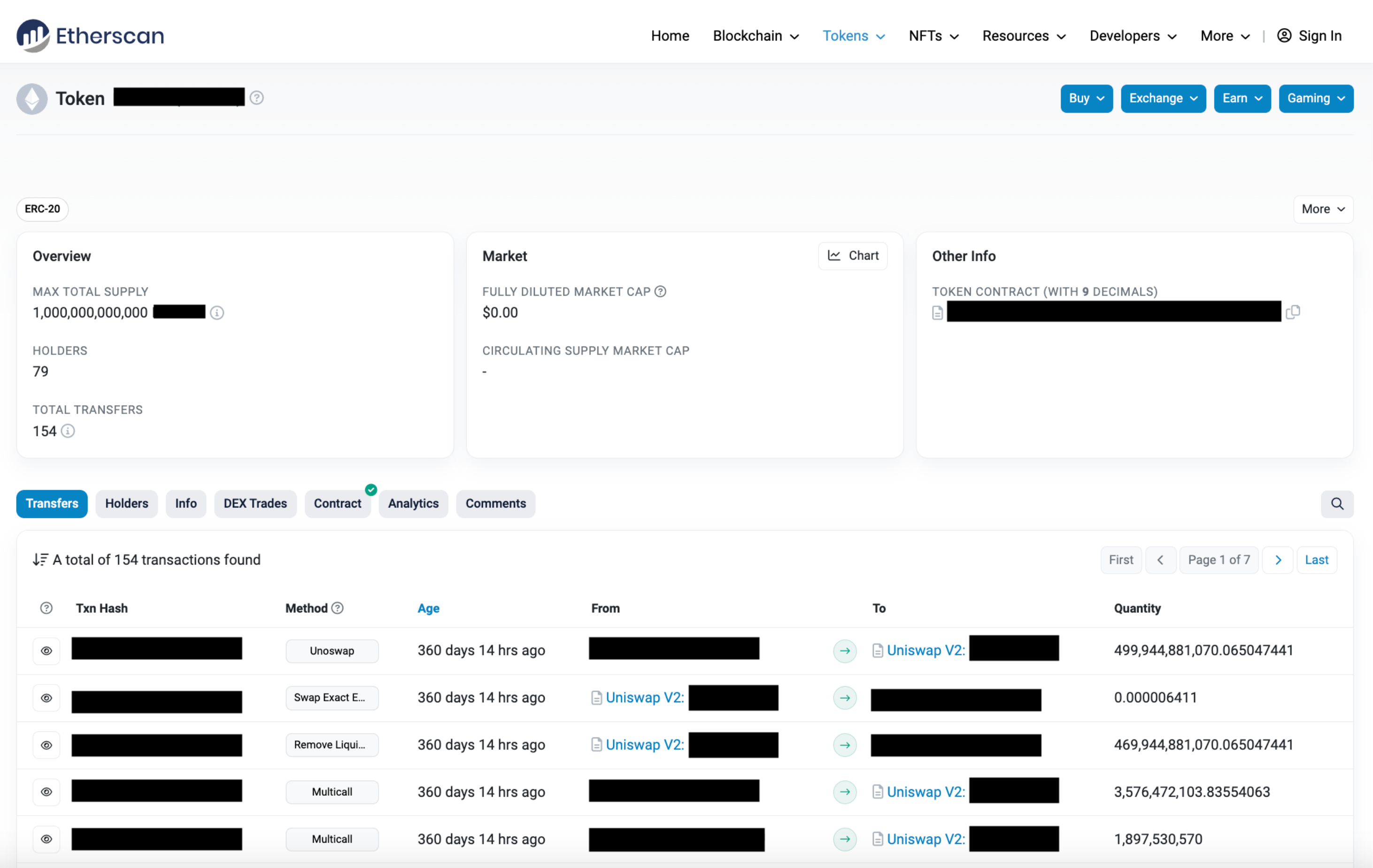}
    \caption{Etherscan token page for Token 1.}
    \label{fig10}
\end{figure*}

\begin{figure*}[h!]
    \centering
    \includegraphics[width=14cm]{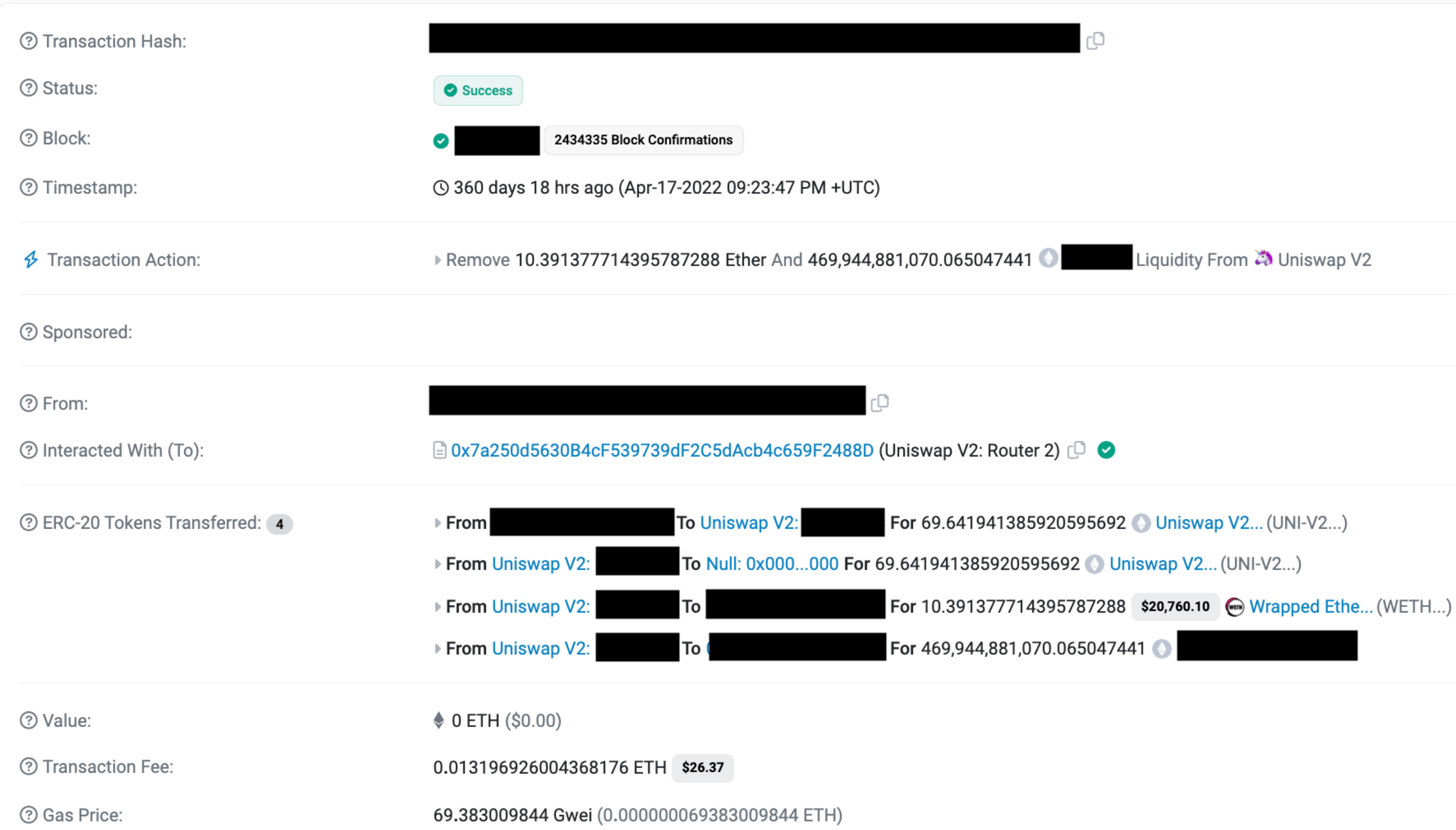}
    \caption{Scammer removing liquidity for Token 1 on Uniswap.}
    \label{fig11}
\end{figure*}

\begin{figure*}[h!]
    \centering
    \includegraphics[width=14cm]{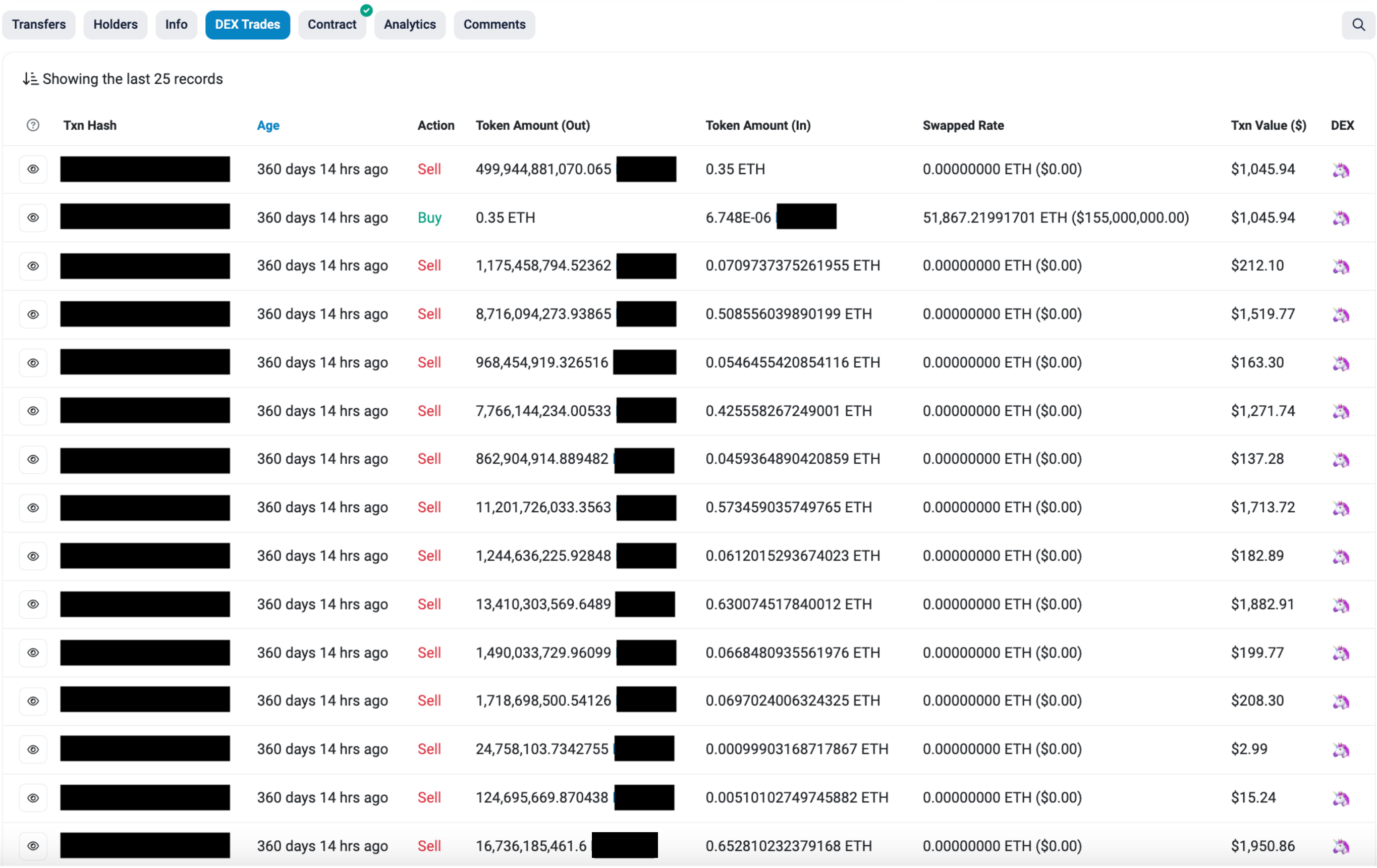}
    \caption{Dex Trades tab on Etherscan for Token 1.}
    \label{fig12}
\end{figure*}

\begin{figure*}[t!]
    \centering
    \includegraphics[width=12cm]{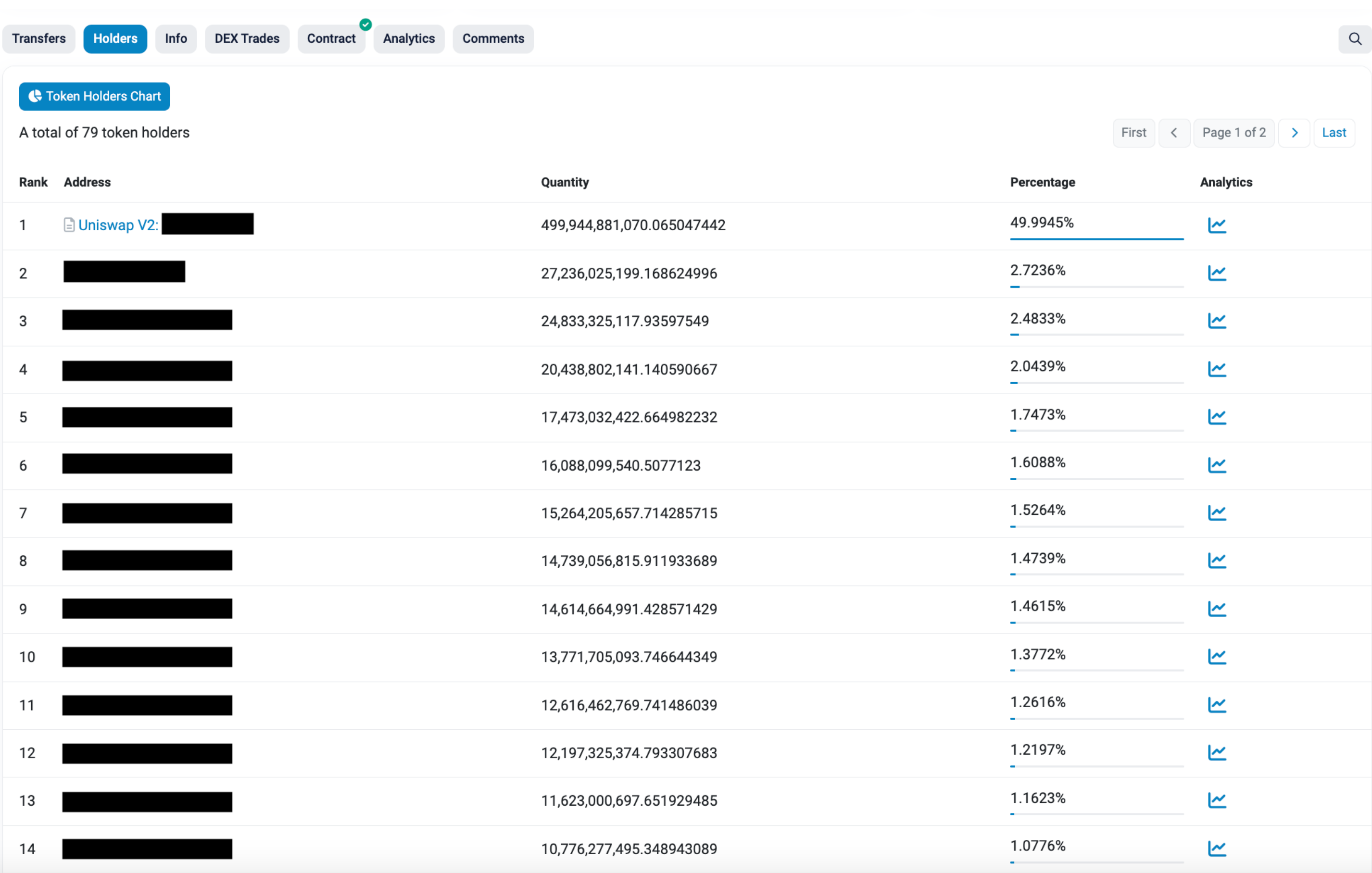}
    \caption{Holders tab on Etherscan for Token 1.}
    \label{fig13}
\end{figure*}

\begin{figure*}[h!]
    \centering
    \includegraphics[width=12cm]{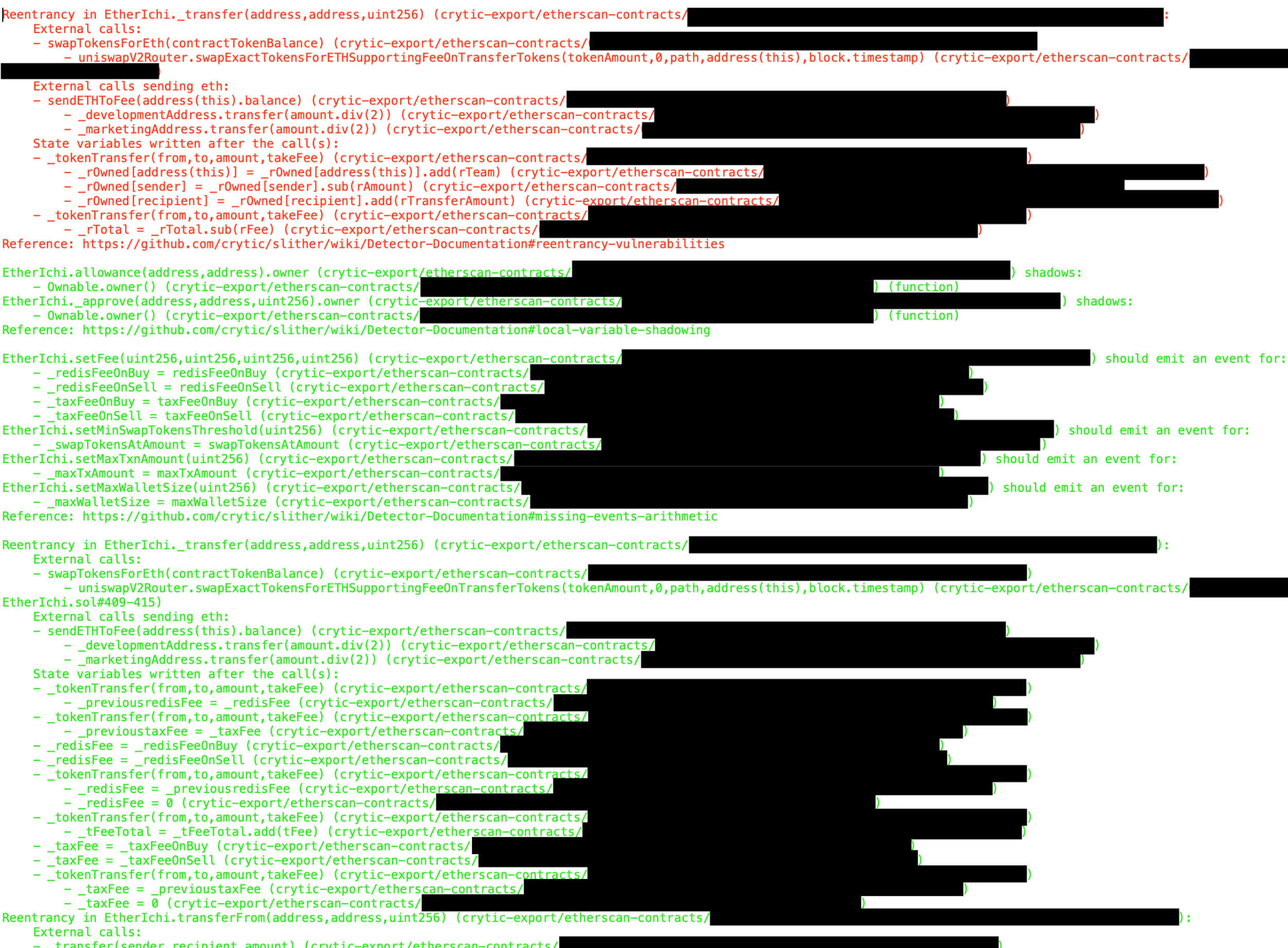}
    \caption{Slither analysis of Token 1 contract.}
    \label{fig14}
\end{figure*}

Figure \ref{fig10} shows the Etherscan token page for Token 1. As discussed in our Method section, we manually examined every ERC-20 token transfer involving Token 1. We focused on identifying token events like adding and removing liquidity and token swaps. 
\\
\\
\\
\\
\\
\\
\\
\\
We do not reproduce every token event or transfer here, however, Figure \ref{fig11} shows an example of one such key event: the perpetrator removing liquidity on Uniswap.
\pagebreak

Figures \ref{fig12} and \ref{fig13} show additional Etherscan tools we used in our investigations, namely, the Dex Trades tab of the Token 1 page (which shows the price of Token 1 for each trade) and the Holders tab (used to identify potential victims left holding Token 1 after the rug pull), respectively. Table \ref{table1} shows the full results of our fraud investigation of Token 1.

Figure \ref{fig14} shows the output from Slither for the analysis of the Token 1 contract. Section 3.1.1 details the outcome of our smart contract analyses.

\begin{figure*}[h!]
    \centering
    \includegraphics[width=14cm]{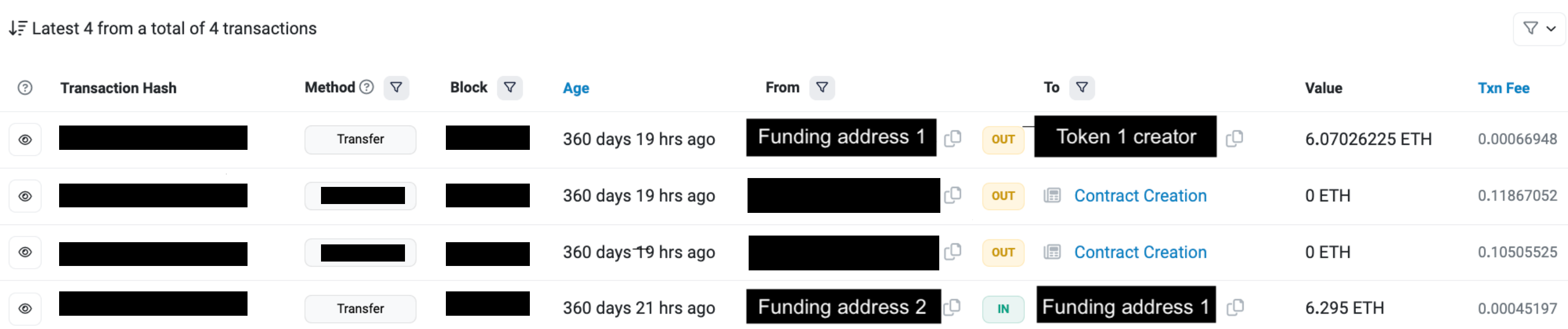}
    \caption{Etherscan depiction of funding scammer wallet.}
    \label{fig15}
\end{figure*}

\begin{figure*}[h!]
    \centering
    \includegraphics[width=14cm]{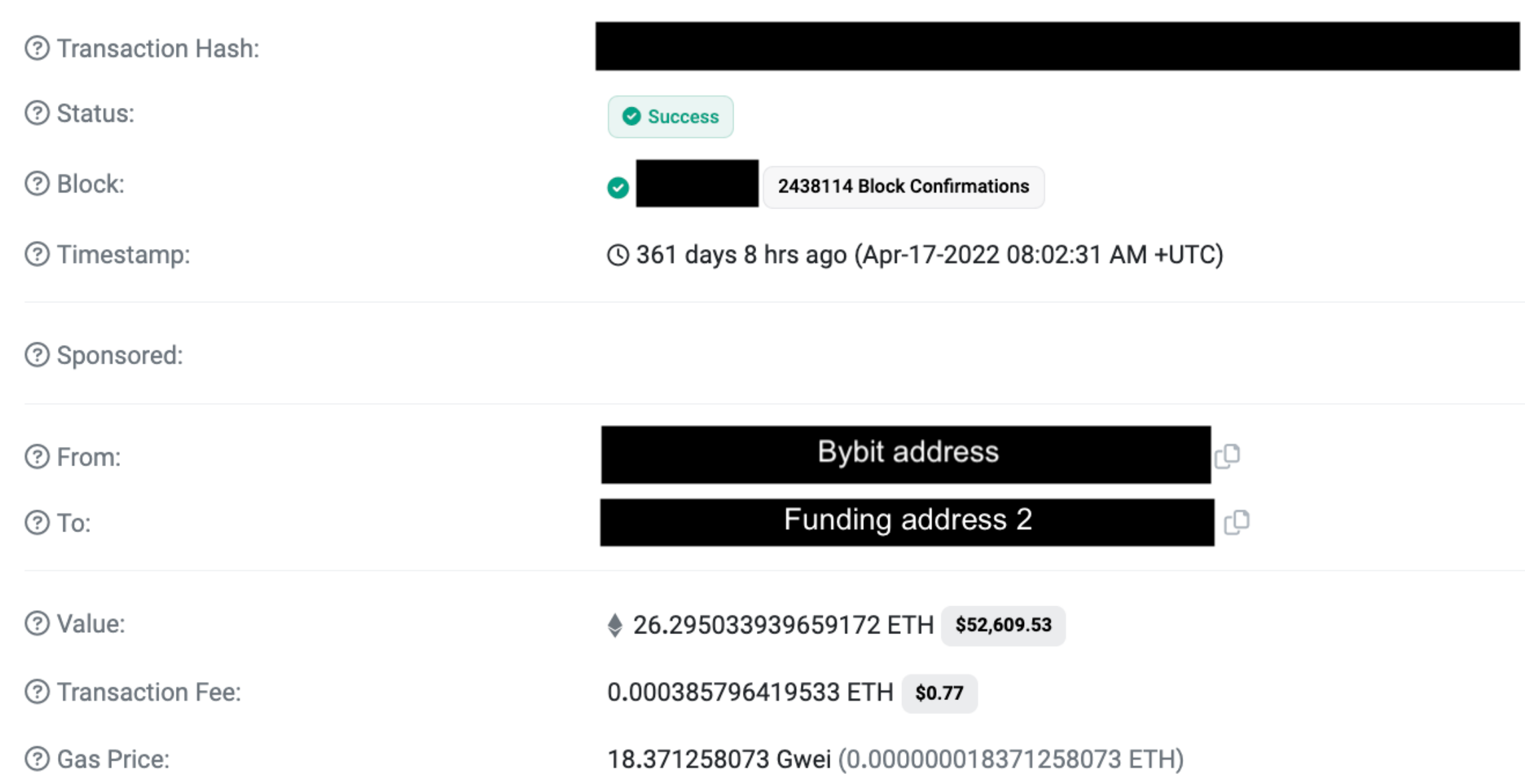}
    \caption{Etherscan depiction of transfer from Bybit to address funding scammer wallet.}
    \label{fig16}
\end{figure*}

\begin{figure*}[p]
    \centering
    \includegraphics[width=8cm]{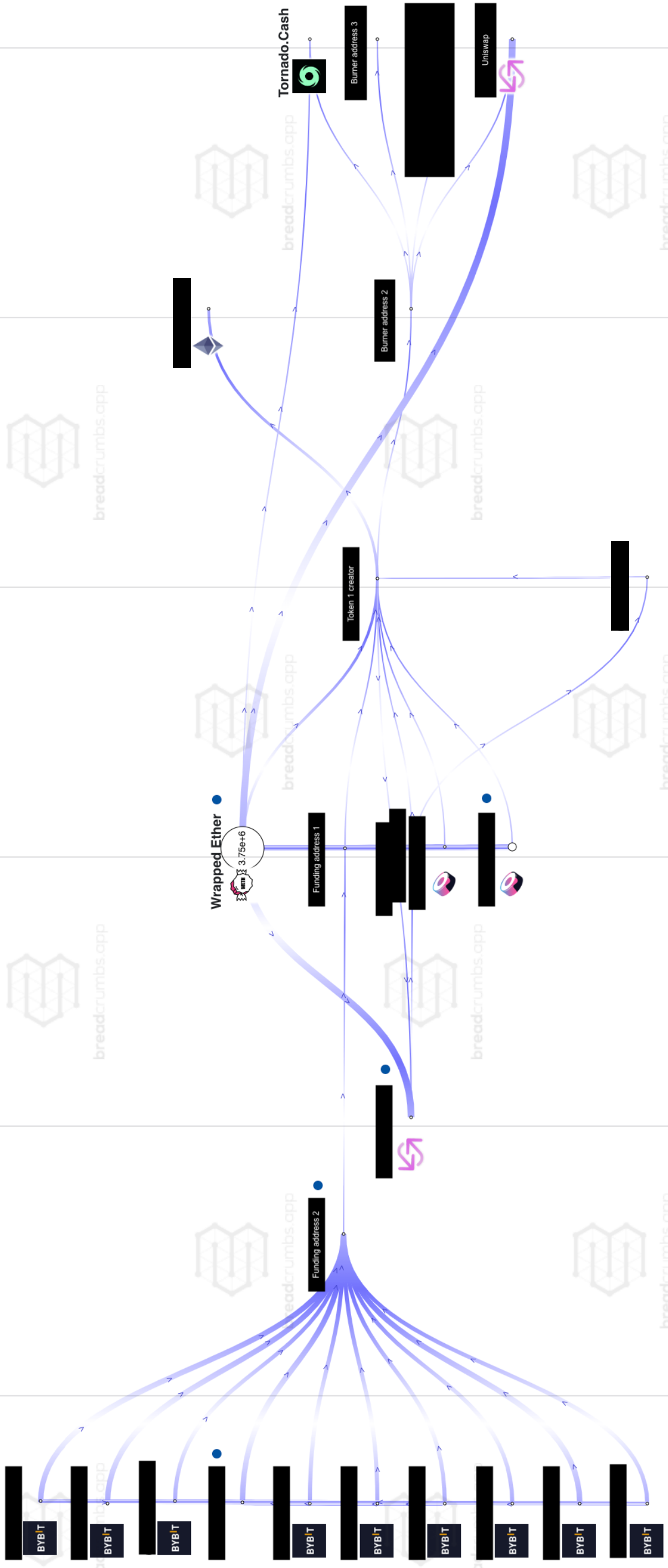}
    \caption{Breadcrumbs depiction of funding scammer wallet and laundering some proceeds of Token 1 scam.}
    \label{fig17}
\end{figure*}

\clearpage
Figures \ref{fig15} and \ref{fig16} show the funding of the scammer address on Etherscan. This is also depicted in Figure \ref{fig17}, which is a selected screenshot from our Breadcrumbs-based money laundering investigation of Token 1. This figure shows the process of funding the scammer wallet and the scammer laundering money through a burner address, then sending some funds to another burner address, Uniswap, and Tornado Cash. The full money laundering scheme is depicted in Figure \ref{fig8}. Further results of these money laundering analyses can be found in Section 3.2 of the body of this article.

\pagebreak
\pagebreak

\bibliographystyle{elsarticle-harv}
\bibliography{cas-refs}



\end{document}